\documentclass[journal]{IEEEtran}
\usepackage{cite}

\ifCLASSINFOpdf
\else
\fi
\usepackage{amsmath}
\usepackage{amsthm}
\ifCLASSOPTIONcompsoc
  \usepackage[caption=false,font=normalsize,labelfont=sf,textfont=sf]{subfig}
\else
  \usepackage[caption=false,font=footnotesize]{subfig}
\fi
\usepackage{amssymb, graphicx, bbm}
\usepackage{xcolor}

\newtheorem{theorem}{Theorem}
\newtheorem{corollary}{Corollary}
\newtheorem{definition}{Definition}
\newtheorem{proposition}{Proposition}

\newtheorem{lemma}{Lemma}

\begin{document}
%
\title{Throughput Maximization in Uncooperative Spectrum Sharing Networks}
%
%
%

\author{Thomas~Stahlbuhk,
        Brooke~Shrader, 
        and~Eytan~Modiano 
\thanks{This work was sponsored by NSF Grants  CNS-1524317 and AST-1547331. 
DISTRIBUTION STATEMENT A. Approved for public release. Distribution is unlimited.

This material is based upon work supported by the United States Air Force under Air Force Contract No. FA8702-15-D-0001. Any opinions, findings, conclusions or recommendations expressed in this material are those of the author(s) and do not necessarily reflect the views of the United States Air Force.}
\thanks{T. Stahlbuhk and B. Shrader are with MIT Lincoln Laboratory, Lexington, MA, 02421. E. Modiano is with the Massachusetts Institute of Technology, Cambridge, MA, 02139. This paper was presented in part in \cite{stahlbuhk}.}}

\maketitle

\begin{abstract}
Throughput-optimal transmission scheduling in wireless networks has been a well considered problem in the literature, and the method for achieving optimality, MaxWeight scheduling, has been known for several decades.  This algorithm achieves optimality by adaptively scheduling transmissions relative to each user's stochastic traffic demands.  To implement the method, users must report their queue backlogs to the network controller and must rapidly respond to the resulting resource allocations.  However, many currently-deployed wireless systems are not able to perform these tasks and instead expect to occupy a fixed assignment of resources.  To accommodate these limitations, adaptive scheduling algorithms need to interactively estimate these uncooperative users' queue backlogs and make scheduling decisions to account for their predicted behavior.  In this work, we address the problem of scheduling with uncooperative legacy systems by developing algorithms to accomplish these tasks.  We begin by formulating the problem of inferring the uncooperative systems' queue backlogs as a partially observable Markov decision process and proceed to show how our resulting learning algorithms can be successfully used in a queue-length-based scheduling policy.  Our theoretical analysis characterizes the throughput-stability region of the network and is verified using simulation results.
\end{abstract}


%
\IEEEpeerreviewmaketitle

\section{Introduction}
%
%
%
%

\IEEEPARstart{T}{o} share their available resources, networked wireless systems must schedule their transmissions to meet the network's throughput demands while avoiding interference.  In the seminal work of \cite{tassiulas} and \cite{tassiulas2}, an adaptive queue-length-based algorithm, MaxWeight scheduling, was shown to achieve throughput-optimality; i.e., for any achievable traffic demand put on the network, MaxWeight successfully schedules the transmissions to meet the demand.  Over the past decades, the MaxWeight algorithm and its extensions have had great success and have been applied to network switching \cite{mckeown}, satellite communications \cite{neely0}, ad-hoc networking \cite{lin, chen2}, packet-delivery-time reduction \cite{joo2}, scheduling with selective and delayed feedback \cite{karaca, deghel}, multicasting/broadcasting \cite{sinha1, sinha2, sinha3}, multi-user MIMO \cite{bethanabhotla}, and age-of-information minimization \cite{kadota1, kadota2}.  The MaxWeight algorithm makes scheduling decisions sequentially-over-time by first observing the backlog of queued packets at each node and then using these observations to adaptively schedule a simultaneous non-conflicting set of links to transmit.  This operation requires two things from the users in the network: the users must be able to report their queue backlogs to the network controller and respond to the resulting schedule.  However, many currently-deployed (legacy) communication systems cannot accommodate these tasks.

Instead, these \emph{uncooperative users} expect to occupy a fixed assignment of channel resources.  For example, the uncooperative users could be multiple access systems, where the signaling dimensions are divided among the users along time, frequency, or code axes \cite{goldsmith}.  Each uncooperative user expects to be assigned a fixed partition, which it has repetitive access to, and the legacy protocols used by the uncooperative users may not accommodate rapid reassignments. Additionally, a common method for achieving low-power communications is to have nodes sleep for long cycles, waking up periodically to communicate if necessary \cite{dementyev, narendra}. If using this technique, the powered down uncooperative users will miss the network controller's directions and cannot waste power engaging in complex control signaling to support transmission scheduling.

In this work, we examine algorithms that can incorporate uncooperative-legacy communication systems into an adaptive-scheduling network.  For those users that cannot report their queue backlogs, the method estimates the backlog at each time-step and makes scheduling decisions accounting for the user's predicted behavior.  We formulate the queue-backlog estimation as an infinite state-space partially observable Markov decision process (POMDP), which is difficult to solve.  We derive upper and lower bounds on the optimal policy for the process and show how to incorporate the results into a queue-length-based adaptive scheduling algorithm.  Importantly, the addition of uncooperative users in the system fundamentally changes the throughput-stability region, which is defined as the set of all traffic demands that can be successfully scheduled by any algorithm.\footnote{The throughput-stability region has also been referred to as the network capacity region, stability region, and stable-throughput region in the literature.}  This region can be viewed as a multi-dimensional area with axes corresponding to requested traffic demands placed on each user.  Our results characterize the geometric-shape of this region and provide a tight scheduling algorithm in certain regimes.  Our theoretical results are verified through simulation.

Our work herein has some relation to cognitive radio networks, which have also been studied using stochastic models \cite{chen0, ahmad, gai1, liu, levorato} and queueing models \cite{urgaonkar, jeon, kompella}.  In cognitive radio, a secondary network gains access to a channel only when it causes minimal conflict with a primary network that owns the resource.  Our problem setup is fundamentally different in that all users are equally important and a scheduler must therefore learn how to manipulate the uncooperative users to achieve throughput-optimality.  Our work is focused on scheduling algorithms that must simultaneously estimate network parameters and make control decisions.  In this direction, learning algorithms for channel assessment and medium access control have been studied in \cite{anadkumar, avner, cayci, combes, gai2, kalathil, krishnasamy, lelarge, liu2, nayyar, stahlbuhk1, tekin, zhange, zhou} under the multi-armed bandit (MAB) framework \cite{bubeck}, and \cite{stahlbuhk2, krishnasamy2} modified the MaxWeight algorithm with a learning component for channel assessment.  In contrast to these works, we consider estimating queue backlogs not channel capacity.

This paper is thematically connected to the work in \cite{jones, rai2}, which applied adaptive algorithms to overlay network routing on legacy networks with unknown queue backlogs.  Our work is also concerned with adaptive network control in the presence of unknown queue backlogs but focuses on transmission scheduling where overcoming channel conflicts is the primary goal.  Finally, the MaxWeight-inspired scheduling policy that we derive belongs to the subclass of policies named Longest Queue First (LQF) that was considered in \cite{dimakis}.  Our proof of the stability of this policy is greatly inspired by \cite{armony, ross} which showed the optimality of MaxWeight algorithms, termed projective cone scheduling in those works, on non-stochastic traffic demands.  We chose this approach for our proofs since the Lyapunov methods used in the proof of MaxWeight \cite{tassiulas, tassiulas2} require bounded second moments, which we do not assume.  As a result, our proof of stability may be of interest outside the narrower context of this work.    

This paper is organized as follows.  In Section~\ref{section_system_description_and_preliminaries}, we describe our network model. We proceed to analyze a network of two users and one channel resource in Section~\ref{section_two_user_network} and derive bounds on the achievable throughput by formulating the problem as a POMDP.  In Section~\ref{section_multiple_user_network}, we use these results to establish necessary and sufficient conditions for the throughput-stability of networks with more than two users and multiple channel resources.  In particular, we formulate the Longest Queue First policy that stabilizes any network that meets our sufficiency conditions.  We verify our theoretical results through simulation. In Section~\ref{channel_assignment}, we address the computational complexity of assigning uncooperative-legacy users to channels.  A subset of this work first appeared in \cite{stahlbuhk}.

\begin{figure}[t!]
\centering
    \includegraphics[width=0.5 \textwidth]{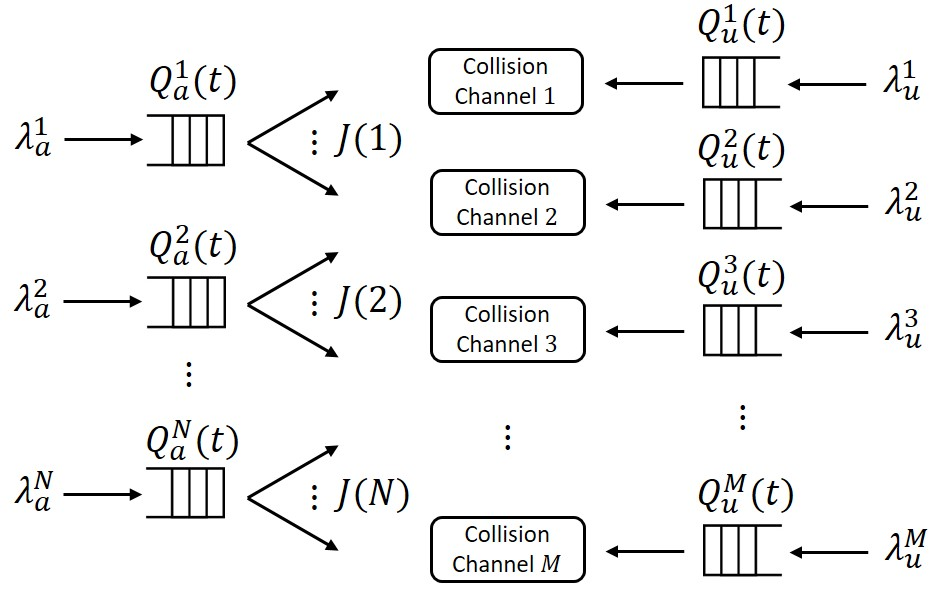}
\caption{Network of $N$ adaptive users (left) and $M$ uncooperative users (right).}
\label{system_mu_illustration}
\end{figure}

 




\section{Problem Setup}
\label{section_system_description_and_preliminaries}

\subsection{Network Composition}
We consider a network consisting of $N$ adaptive users and $M$ uncooperative users operating over time slots $t = 0, 1, 2, \dots$.  The users communicate to a common receiver (e.g., a base station or access point) and the network is operated by a centralized controller (i.e., scheduler).  See Fig.~\ref{system_mu_illustration}.  At each time step, the adaptive users report their queue backlogs to the controller and respond to the controller's transmission decisions, while the uncooperative users do neither.  Each uncooperative user $j \in \{1, 2, \dots, M\}$ has its own assigned channel resource (often simply referred to as ``channel" henceforth).  This channel resource could correspond to a time slot in a repeating TDMA frame, a frequency allocation, or a unique frequency-hopping pattern.  For now, we will assume the channel resource has been assigned to the uncooperative user at the start of time by some unspecified method and address the assignment problem in Section~\ref{channel_assignment}.

The uncooperative users do not have their transmissions scheduled by the controller.  Instead, in each time slot that its queue is nonempty, an uncooperative user transmits one packet on its assigned channel.  Packets arrive independently to each uncooperative users' queue as a Bernoulli processes with rates $\lambda^j_u$ and can be transmitted in the time slot in which they arrive.  Let $A^j_u(t) \in \{0, 1\}$ be the number of arrivals to user $j$ at time slot $t$ and $Q^j_u(t)$ be the number of packets in queue after arrival $A^j_u(t)$.  Under our model, an empty channel resource with no assigned uncooperative user is equivalent to a channel with an uncooperative user having arrival rate $\lambda^j_u = 0$ (i.e., having a user that never needs to transmit).

The adaptive users also communicate by transmitting packets over the channel resources.  Each adaptive user $i \in \{1, 2, \dots, N\}$ has a subset of channels $J(i) \subseteq \{1, 2, \dots, M\}$ on which it is allowed to transmit, which may be restricted because of regulatory constraints or the limitations of its radio frequency hardware.  At each time step, each user $i$ can be scheduled by the controller to transmit on any subset of $J(i)$, transmitting one packet on each channel in the subset, subject to the constraint that only one adaptive user may transmit on a channel at a given time.  Packets arrive to the queue of user $i$ according to an \emph{i.i.d.} stochastic process $A^i_a(t)$ with rate $\lambda^i_a$ and can be transmitted in the time slot in which they arrive. We use $Q^i_a(t)$ to denote the number of packets in queue after arrivals $A^i_a(t)$.  We assume $A^i_a(t) \leq A_{max}$ for some finite constant and all arrival processes to the network are independent.

\subsection{Channel Behavior and Receiver Feedback}
Each channel resource is a \emph{collision channel}.  Therefore, for each channel, if only one user transmits during a time slot, the transmission is successful and the packet departs that user's queue, but if multiple users transmit on the channel during the time slot, the packets collide, all transmissions fail, and the packets remain in their respective queues awaiting future successful transmission.  Because each uncooperative user exclusively transmits on its dedicated channel resource and adaptive users are centrally controlled, collisions can only occur between uncooperative and adaptive users.

At the end of each time slot $t$, the network controller receives ternary feedback indicating whether each channel contained a successful transmission, a collision, or was idle.  We assume that the controller knows the arrival processes to the uncooperative users are independent Bernoulli and knows the values of $\lambda^i_a$ and $\lambda^j_u$ for all $i$ and $j$ (this last assumption will be addressed in Subsection~\ref{subsection_longest_queue_first}).  However, the key challenge facing the controller is that it cannot either observe the uncooperative users' arrival processes $A^j_u(t)$ or queue backlogs $Q^j_u(t)$ directly.  Instead, it must rely upon its history of previous actions, the queue backlogs of the adaptive users, and the ternary feedback to determine whether to schedule an adaptive user to transmit on each channel resource.

Before continuing, we establish some brief notation and definitions.  At each time $t$, we denote the number of packets to depart the queues of uncooperative user $j$ and adaptive user $i$ with $\tilde{B}^j_u(t) \in \{0, 1\}$ and $\tilde{B}^i_a(t) \in \left\{0, 1, \dots, \left\lvert J(i) \right\rvert \right\}$, respectively.  Moreover, without loss of generality, we assume at time $t = -1$, $Q^j_u(-1) = 0$ and $Q^i_a(-1) = 0$ for all $i$ and $j$.  Thus, we may write
\begin{equation}
\label{secondary_user_queue_size_summation}
Q^i_a(t) = \sum_{\tau=0}^t A^i_a(\tau) - \sum_{\tau=0}^{t-1} \tilde{B}^i_a(\tau)
\end{equation}
with an analogous equation holding for uncooperative user $j$.

We will find it convenient to sometimes indicate over which channel a packet was successfully transmitted.  Therefore, at time $t$, we let $\tilde{B}^{(i,j)}_a(t) \in \{0,1\}$ indicate the number of packets that departed the queue of user $i$ because of a successful transmission on channel $j$, and we let $\tilde{B}^j(t) \in \{0, 1\}$ indicate the total number of packets that departed over the channel.  Clearly, $\tilde{B}^i_a(t) = \sum_{j \in J(i)} \tilde{B}^{(i,j)}_a(t)$ and $\tilde{B}^j(t) = \sum_{i : j \in J(i)}  \tilde{B}^{(i,j)}_a(t)$.

Furthermore, we allow adaptive users to transmit dummy packets, which are transmissions that do not service a packet in the user's queue (i.e., transmissions that do not carry data with them).  Dummy packet transmissions are capable of colliding with uncooperative user transmissions, and importantly, they allow adaptive users to interact with a channel even when their queues are empty.  We use $B^i_a(t) \in \left\{0, 1, \dots, \left\lvert J(i) \right\rvert \right\}$, $B^{(i,j)}_a(t) \in \{0,1\}$ and $B^j(t) \in \{0, 1\}$ (analogous to the above) to indicate successful transmissions including dummy packets.

\subsection{Queue Stability}
Our focus in this work will be on developing scheduling algorithms that stablize all users' queues.  We now give several definitions of stability that are widely used in the literature \cite{neely}.

\begin{definition}
\label{definition_of_mean_rate_stability}
The queue of user $i$ is mean rate stable if and only if
\begin{equation*}
\lim_{T \to \infty} \frac{E\left[ Q^i_a(T) \right]}{T} = 0.
\end{equation*}
\end{definition}
\begin{definition}
\label{definition_of_rate_stability}
The queue of user $i$ is rate stable if and only if
\begin{equation*}
\lim_{T \to \infty} \frac{Q^i_a(T)}{T} = 0, \mbox{\quad with prob. $1$}.
\end{equation*}
\end{definition}
The definitions apply similarly to uncooperative users.  Note that in this work, because the number of arrivals to any queue is bounded above by a finite constant, by \cite[Theorem 1]{neely} the rate stability of a queue will imply that it is mean rate stable as well.  Therefore, mean rate stability is the weaker definition. We define the network to be stable if all of its queues are rate stable. The throughput-stability region is the set of all arrival rates that could be placed on the network such that any scheduling algorithm could achieve network stability.

\begin{definition}[Throughput Stability Region]
The throughput stability region,
\begin{multline*}
\Lambda \triangleq \Bigg\{ \lambda^i_a, \lambda^j_u : \mbox{ for some scheduling policy, with prob. $1$, } \\ \lim_{T \to \infty} \frac{Q^i_a(T)}{T} = 0, \lim_{T \to \infty} \frac{Q^i_p(T)}{T} = 0, \forall i \in [N], j \in [M] \Bigg\}
\end{multline*}
\end{definition}
Definitions~\ref{definition_of_mean_rate_stability}~and~\ref{definition_of_rate_stability} imply that the rate of departures from the queue equals the rate of arrivals in expectation and almost surely, respectively \cite{neely}.  Throughout this work, we will refer to the service rate to a user's queue as its \emph{throughput}.  The network is therefore stable if each user has a throughput that matches its arrival rate.

\subsection{Objective}

\begin{figure}
  \centering
  \includegraphics[width=0.5 \textwidth]{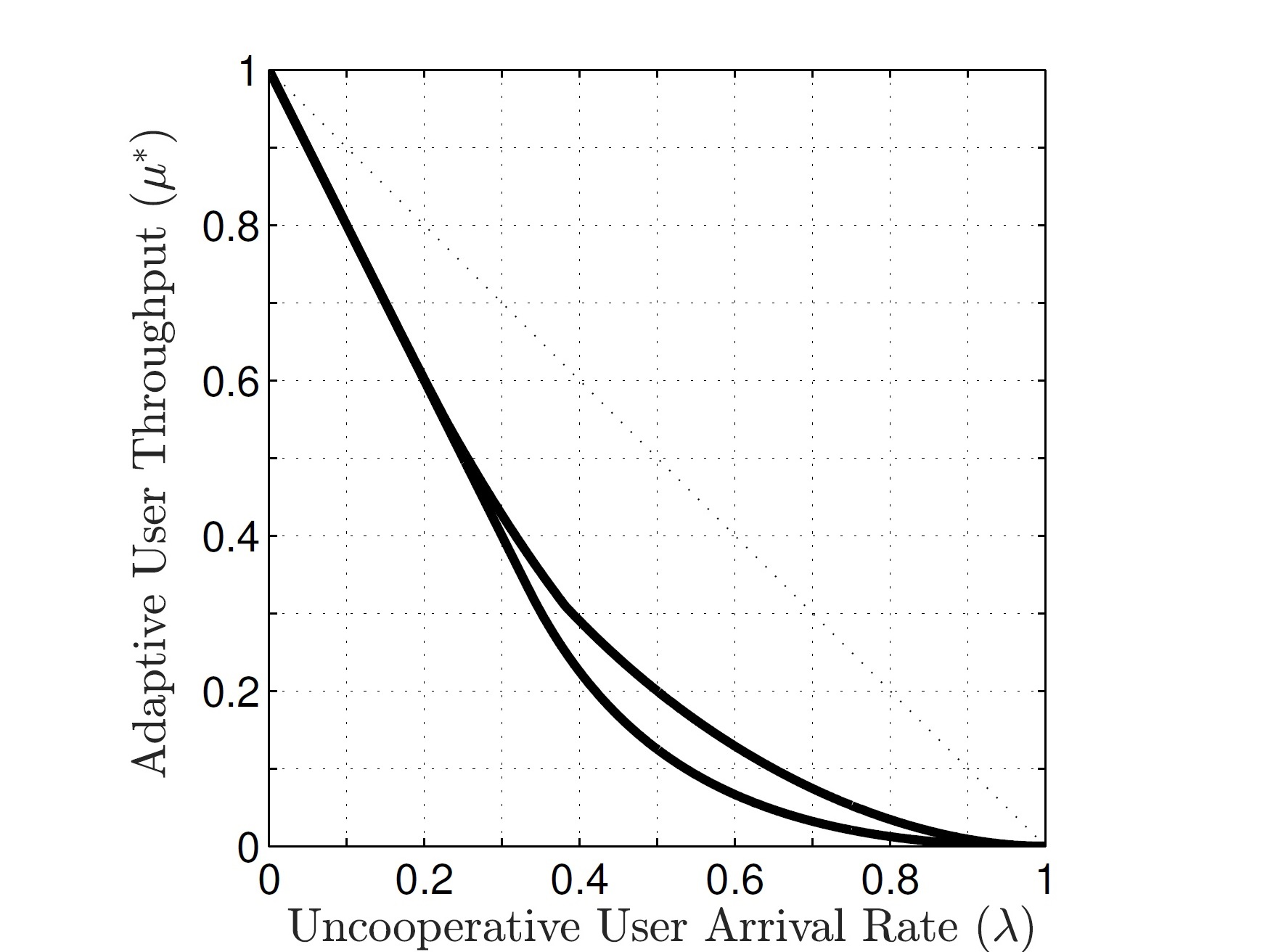}
  \caption{The upper and lower bounds on $\mu^*$ (the attainable throughput of the adaptive user) versus $\lambda$ (the throughput of the uncooperative user) on a single channel.  The bounds are shown in dark solid lines.  The line $1 - \lambda$ is shown for reference in light dots.}
  \label{bounds_plot}
\end{figure}

We are concerned with establishing scheduling policies that rate stabilize all queues in the network.  Note that an instantiation of the above problem is completely parameterized by the set of variables $\lambda^i_a$ and $\lambda^j_u$ as well as subsets $J(i)$.  For example, the throughput-stability region for a network consisting of two adaptive users sharing a common channel (without an uncooperative user on the channel) is given by $\lambda^1_a + \lambda^2_a \leq 1$.  This region consists of all arrival rates to the two users that can be stabilized by any policy and arises from the controller's need to time share the channel.  Note that the region is the convex hull of scheduling all time to either one user or the other.  This convex polytope shape is typical for time-sharing problems \cite{neely_book}.  In contrast, in Fig.~\ref{bounds_plot}, we show upper and lower bounds on the stability region of a network consisting of one adaptable and one uncooperative user.  Note that as opposed to the region for two adaptive users, the region in Fig.~\ref{bounds_plot} does not appear to be a polytope nor is it convex.  This structure arises from the partial observability of the problem, and the controller's varying ability to provide accurate estimates of the uncooperative user's queue backlog for different arrival rates.


In the following, we derive necessary and sufficient conditions on the throughput-stability region of networks with adaptable and uncooperative users.  Our proof of sufficiency is constructive and will define implementable policies. We proceed in our analysis by first examining a network consisting of one adaptive user and one uncooperative user in Section~\ref{section_two_user_network}.  Our examination will establish lower and upper bounds for the stability region of this network.




\section{Two User Network with One Channel Resource}
\label{section_two_user_network}

In this section, we examine the throughput-stability region of a network consisting of one adaptive user and one uncooperative user.  Since there are only two users in the network we will simplify our notation for this section.  Namely, we will let $A(t) \in \{0, 1\}$ denote the Bernoulli arrival process to the uncooperative user, $\lambda \in (0, 1)$ denote the rate of this process, and $Q(t)$ denote the queue backlog of the uncooperative user after arrival $A(t)$.\footnote{The case $\lambda = 0$ (the uncooperative user never transmits) and $\lambda = 1$ (always transmits) clearly gives the adaptive user a throughput of $1$ and $0$, respectively.}

We proceed to derive lower and upper bounds on the maximum rate at which the adaptive user can make successful transmissions as a function of $\lambda$, the packet arrival rate to the uncooperative user (see Fig.~\ref{bounds_plot}).\footnote{Our focus is on characterizing the maximum rate that the adaptive user can successfully access the channel without colliding with the uncooperative user.  Therefore, the adaptive user will transmit dummy packets whenever its queue is empty.}  Importantly, since the users transmit over a collision channel model and the uncooperative user will transmit whenever its queue is nonempty, the adaptive user can only obtain successful transmissions when the uncooperative user's queue is empty.  Therefore, in order to have a nonzero adaptive user throughput, the controller must keep the uncooperative user's queue backlog stable.  If it does not, the uncooperative user's queue backlog will grow to infinity and the adaptive user will never have an opportunity to obtain a successful transmission again.

\subsection{POMDP Model}
\label{POMDP_section}

We proceed to give a concrete formulation of the problem as a POMDP following the notation of \cite{bertsekas_vol1}. At each discrete time $t$, the state of the system is the queue backlog of the uncooperative user, $Q(t) \in \{0, 1, 2, \dots\}$, and the adaptive user may take an action $u(t) \in \mathcal{U} = \left\{TR, NT\right\}$ where $TR$ denotes transmission and $NT$ no transmission.  The state $Q(t)$ is unobservable to the controller.  Instead, ternary feedback provides an observation $z(t) \in \mathcal{Z} = \left\{S, C, I\right\}$ where $S$ denotes a successful transmission, $C$ a collision, and $I$ an idle slot.  The observed ternary feedback is given by the following function of $Q(t)$ and $u(t)$

\begin{equation*}
z(t) = \left\{
\begin{array}{ll}
S, &\mbox{if $Q(t) = 0$ and $u(t) = TR$} \\
S, &\mbox{if $Q(t) > 0$ and $u(t) = NT$} \\
C, &\mbox{if $Q(t) > 0$ and $u(t) = TR$} \\
I, &\mbox{if $Q(t) = 0$ and $u(t) = NT$}
\end{array}
\right..
\end{equation*}
To inform its decisions, the controller has at its disposal the history $H(t)$ of all past actions and observations (i.e., $H(t) = \left(u(0), \dots, u(t-1),z(0), \dots, z(t-1)\right)$).  Without loss of generality, we assume $Q(-1)=0$ and this fact is known to the controller.

A solution to the POMDP is a policy $\pi = f_0, f_1, f_2, \dots$ that provides for each time $t$ a mapping $f_t$ such that $f_t\left(H(t)\right) \in \mathcal{U}$.  Define by $B(t)$ the reward process of the POMDP where
\begin{equation*}
B(t) = \left\{
\begin{array}{ll}
1, \mbox{\quad if $Q(t) = 0$, $u(t) = TR$} \\
0, \mbox{\quad otherwise}
\end{array}
\right..
\end{equation*}
(i.e., a reward of $1$ is accrued for each successful adaptive user transmission.)  An optimal solution to the POMDP is a policy that achieves
\begin{equation}
\mu^* \triangleq \underset{\pi}{\max} \text{ } \underset{T \to \infty}{\lim \sup} \text{ } E\left[\frac{1}{T} \sum_{t=0}^{T-1} B(t)\right].
\label{objective}
\end{equation}
This objective is equivalent to maximizing the throughput of the adaptive user given that at each time $t$ the adaptive user always has an available packet to send.

Given $H(t)$, a probability mass function on the distribution of $Q(t)$ may be computed.  In general, evaluating the optimal solution to a POMDP is difficult \cite{bertsekas_vol1}.  In the following sections we derive lower and upper bounds on \eqref{objective} as a function of $\lambda$.  Note that since successful transmissions by the adaptive user can only occur during time slots when the uncooperative user's queue is empty, our POMDP formulation implicitly requires that the uncooperative user achieves a throughput of $\lambda$ (which is equal to its arrival rate).

\subsection{Lower Bound on the Adaptive User's Throughput}
\label{lower_bound_section}
We derive a lower bound on \eqref{objective} by evaluating the performance of a class of simple, suboptimal randomized stationary policies, $\pi^{lb}$.  Policy $\pi^{lb}$ is defined as follows: at each time slot $t$, $\pi^{lb}$ does not schedule the adaptive user to transmit if the ternary feedback indicates a collision at time slot $t-1$ (i.e., back off after collisions) and otherwise schedules a transmission with probability $p$.  For a given value of $\lambda$, $p$ can be optimized to obtain the maximum adaptive user throughput attainable by this class of policies.  This gives the following.

\begin{theorem}
\label{lower_bound_thm}
There exists a value of $p$, denoted $p^*$, such that the randomized stationary policy $\pi^{lb}$ achieves
\begin{equation}
\mu^*_{lb} \triangleq \underset{T \to \infty}{\lim} E\left[\frac{1}{T} \sum_{t=0}^{T-1} B(t) \right] = \left \{
\begin{array}{l}
1 - 2\lambda, \mbox{ for $\lambda \leq \frac{1}{3}$ } \\
\\
\frac{(1-\lambda)^2}{4\lambda}, \mbox{ for $\lambda > \frac{1}{3}$ }
\end{array}
\right.
\label{lower_bound}
\end{equation}
\end{theorem}
The value of $p^*$ attaining \eqref{lower_bound} is given by
\begin{equation}
\label{p_star}
p^* = \left\{
\begin{array}{l}
1, \mbox{ for $\lambda \leq \frac{1}{3}$ } \\
\\
\frac{1-\lambda}{2\lambda}, \mbox{ for $\lambda > \frac{1}{3}$}
\end{array}.
\right.
\end{equation}
We plot \eqref{lower_bound} as a function of $\lambda$ in Fig.~\ref{bounds_plot}.  The proof is based on a simple Markov chain analysis of the policy.  The Markov chain consists of states that indicate the current value of $Q(t)$ and whether the controller is backing off because of a collision in the previous time slot.  Solving for the steady state distribution of the chain and optimizing over $p$ gives the result.  Furthermore, it is easy to show that under the policy
\begin{equation}
\label{lower_bound_wp1}
\lim_{T \to \infty} \frac{1}{T} \sum_{t = 0}^{T-1} B(t) = \mu^*_{lb}, \mbox{\quad with prob. $1$},
\end{equation}
which will be important to later proofs. From this analysis we also find that under policy $\pi^{lb}$, the uncooperative user's queue satisfies the following notion of stability, which implies rate stability (see \cite[Theorem 4]{neely}).
\begin{corollary}
\label{primary_user_bounded_expectation}
Under policy $\pi^{lb}$ with $p = p^*$, 
\begin{equation*}
\underset{T \to \infty}{\lim} E\left[Q(T)\right] < \infty.
\end{equation*}
\end{corollary}

As a final note, the policy we constructed can schedule the adaptive user to transmit when its queue is empty.  This decision will simplify analysis later on and does not impact the throughput-stability region of the network.

\subsection{Upper Bound on the Adaptive User's Throughput}
\label{upper_bound_section}

We next provide an upper bound on \eqref{objective}.  We begin with the following two lemmas.
\begin{lemma}
\label{optimal_1}
A policy that minimizes the expected time between successful adaptive user transmissions, maximizes the adaptive user's throughput.
\end{lemma}
\begin{IEEEproof}
Recall that we assume that at time $t=-1$, $Q(-1) = 0$, and this fact is known by the controller.  Now, assume at time $t-1$, the adaptive user makes a successful transmission.  Then, $H(t) = (u(0), \dots, u(t-1)=TR, z(0), \dots, z(t-1) = S)$ and the adaptive user knows $Q(t-1) = 0$.  However, conditioned on $Q(t-1) = 0$, the actions $u(0), \dots, u(t-1)$ and observations $z(0), \dots, z(t-1)$ are independent of all future events.  Additionally, the future starting at time $t$ is statistically the same as at time $0$.  Therefore, every successful transmission by the adaptive user renews the system and there exists an optimal policy that, after each successful transmission, ignores all actions and observations that preceded the successful transmission.

Under this optimal policy, the intervals between successful transmissions are independent and identically distributed and thus form a renewal process.  By the elementary renewal theorem \cite[Theorem~5.6.2]{gallager} we have
\begin{equation}
\label{elementary_renewal}
\underset{T \to \infty}{\lim} \text{ } E\left[ \frac{1}{T}\sum_{t=0}^{T-1} B(t)\right] = \frac{1}{E\left[D\right]}.
\end{equation}
where $E\left[D\right]$ is the mean time between successful transmissions.  Thus, a policy that minimizes $E\left[D\right]$ must maximize throughput (cf. \eqref{objective}).
\end{IEEEproof}

Before proceeding with the upper bound we will need the following additional lemma.

\begin{lemma}
\label{optimal_2}
There exists a policy that maximizes the adaptive user's throughput and adopts the following rule: if $z(t-1) = C$, then action $u(t) = NT$ is selected.
\end{lemma}

The intuition behind Lemma~\ref{optimal_2} is simple. Since after a collision we know that the uncooperative user's queue is nonempty, a transmission by the adaptive user can only result in another collision and the controller should therefore back off at least one time slot before attempting another transmission.  The proof is omitted for brevity but can be found in \cite[Chapter 3.A.1]{stahlbuhk_thesis}.


\subsubsection{Augmented System Model}

Define $W(t)$ to be the maximum number of packets that could potentially be in the uncooperative user's queue at time slot $t$ (i.e., the maximum possible value of $Q(t)$ given $H(t)$).  Note that every time we observe $z(t-1)=I$ or obtain a successful adaptive user transmission, we know $W(t) = 1$ since either event implies $Q(t-1) = 0$.

Now, to derive an upper bound on the adaptive user's throughput, we augment the observation space $\mathcal{Z}$ to include additional information beyond ternary feedback.  Any policy that ignores this additional feedback is admissible under the original system.  Thus, the optimal solution to the augmented system is an upper bound on the original system.

\begin{figure}
  \centering
    \includegraphics[width=0.4\textwidth]{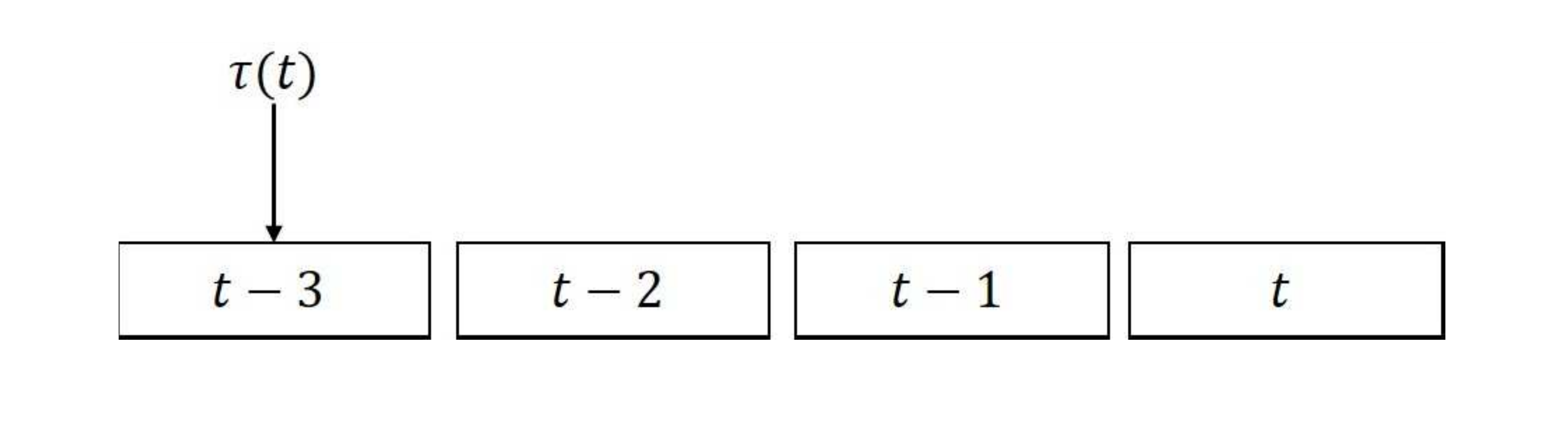}
    \caption{Suppose at time $t-1$ the controller observes an uncooperative user packet with timestamp $t-3$. Then, all uncooperative user packets that arrived up to time $t-3$ must have already been serviced. At time $t$, the controller is only uncertain whether packet arrivals occurred during times $t-2$, $t-1$, and $t$.}
    \label{tau_picture}
\end{figure}

Our augmentation is as follows.  Assume the uncooperative user serves packets in first come first served (FCFS) order and each packet is timestamped with the time slot in which it arrived to the uncooperative user's queue.  Under our augmented system, every time the uncooperative user obtains a successful transmission, the timestamp on the successfully transmitted packet is revealed to the controller.  Thus, the new observation space consists of $\mathcal{Z}$ and the timestamps on the uncooperative user's successfully transmitted packets.  Upon observing a new timestamp the controller knows that all packets that may have arrived to the uncooperative user's queue prior to the timestamp have been successfully transmitted. Define $\tau(t)$ to be the most recently observed timestamp by the start of time slot $t$. Then, uncertainty only remains over those packets that could have arrived to the uncooperative user since time $\tau(t)$.  Clearly, following the observation of a new timestamp, $W(t) = t - \tau(t)$ and this is equal to the number of time slots where the controller has yet to determine the outcome of process $A(t)$ (see Fig.~\ref{tau_picture} for illustration).

\subsubsection{Augmented System as a Total Cost Problem}

We now formulate a stochastic shortest path (SSP) problem with the goal of minimizing the expected time until the next successful transmission by the adaptive user.  Without loss of generality, we analyze the stochastic shortest path problem starting at time $0$, until the first successful adaptive user transmission.  Note that after each successful adaptive user transmission, the controller knows the uncooperative user's queue backlog returned to $0$. Thus, the problem of obtaining the next successful adaptive user transmission starting at a time following a successful transmission is statistically the same as it was at time $0$ (i.e., the problem renews).  We can therefore reapply our policy iteratively, starting over after each successful adaptive user transmission, to obtain the next successful transmission.   By Lemma~\ref{optimal_1}, finding a policy that minimizes the time between successful adaptive user transmissions is equivalent to finding a policy that maximizes the throughput.  

Now, by Lemma~\ref{optimal_2}, there exists an optimal policy that, given a collision at time $t-1$, chooses not to transmit at time $t$; thereby allowing the previously collided uncooperative user packet to be successfully transmitted.  We therefore restrict our attention to policies that take action $u(t) = NT$ whenever $z(t-1) = C$.  In the following, we analyze the augmented system model in order to bound the optimal solution to the original system.

\begin{figure}
  \centering
    \includegraphics[width=0.4\textwidth]{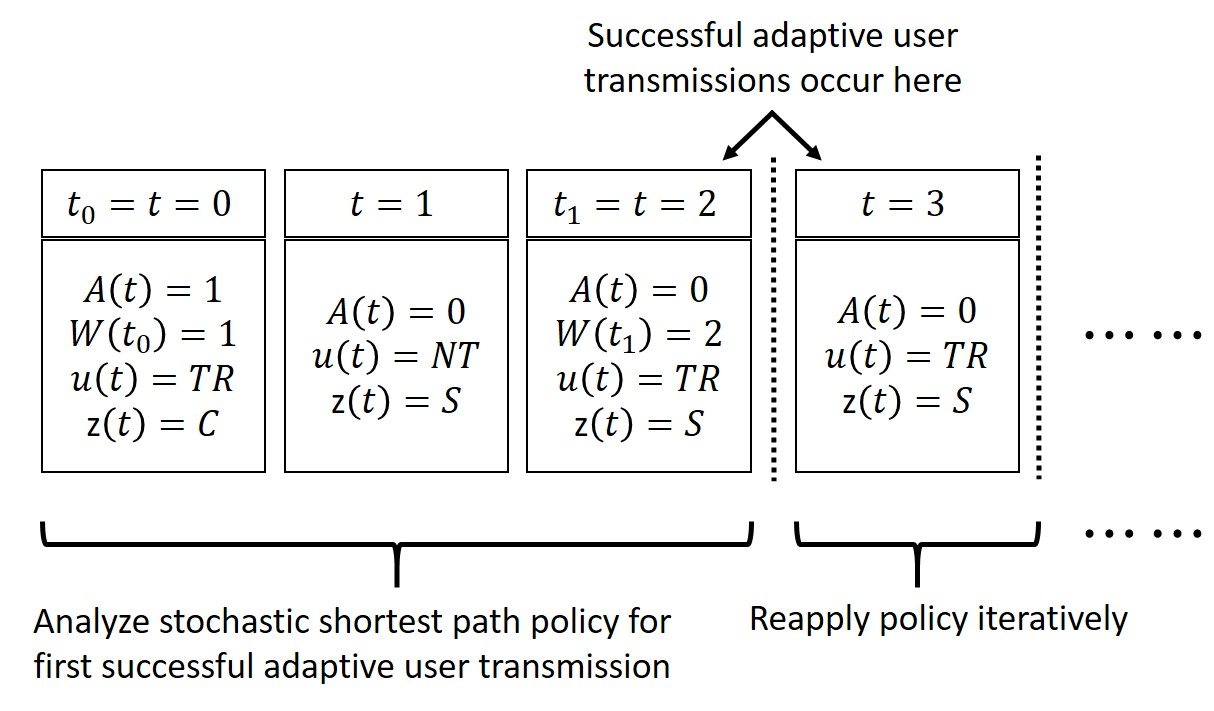}
    \caption{The SSP problem is analyzed starting at time $0$ until the first successful adaptive user transmission.  The subsequence $t_k$ is the time slots not following a collision when transmission decisions are made using $W(t_k)$. In this example, the first successful adaptive user transmission occurs after three time slots.  The policy is reapplied after each successful adaptive user transmission when the SSP renews, which is indicated by dashed lines. }
    \label{time_slots_figure}
\end{figure}

Given the above, we formulate the stochastic shortest path problem as an infinite horizon Markov decision process (MDP) over decision-stage index $k = 0, 1, \dots$ (See Fig.~\ref{time_slots_figure}).  The index $k$ defines a subsequence $t_k$ of the sequence $t$ (starting at $t_0=t=0$) where the subsequence corresponds to the time slots $t$ that do not follow a collision.\footnote{At the start of time we are not backing off from a collision and thus $t_0 = 0$.}  Given that we force the policy to choose $u(t) = NT$ whenever $z(t-1) = C$, one can see that $t_k$ corresponds to those time slots where decisions about which action to take must be made.  Note that $W(t_k)$ completely characterizes the controller's knowledge about $Q(t_k)$ at time $t_k$. Moreover, $P\left( Q(t_k) = q | W(t_k) = w \right) = \left( \substack{w \\ q} \right) \lambda^q (1-\lambda)^{w - q}$. Therefore, an optimal policy can be a function of $W(t_k)$ instead of $H(t_k)$ \cite{bertsekas_vol1}.

We define the states of our stochastic shortest path problem as $X(t_k) \in \mathcal{N} = \{0, 1, 2, \dots\}$ and actions as $u(t_k) \in \mathcal{U}$.  Our state space $\mathcal{N}$ is composed of two parts.  Prior to obtaining a successful transmission, the system is in states $\{1, 2, \dots\}$ which corresponds to the value of $W(t_k)$, the maximum possible size of $Q(t_k)$ given our observations, as defined above.  However, when the adaptive user obtains a successful transmission, the system enters a trapping state, $0$, and remains there for all future indices at no further cost.  Note that state $0$ is the destination state in our stochastic shortest path problem, and we want to reach it with minimum incurred cost.  Once we enter state $0$, the stochastic shortest path problem effectively ends.  We now give the state transition probabilities and (negative) reward function of the MDP.  They are subsequently explained.

For $x \geq 1$, the transition probabilities are given by
\iftrue
\begin{multline}
P\left(X(t_{k+1})=y|X(t_k) = x, u(t_k)\right) \\=
\left\{
\begin{array}{lll}
(1-\lambda)^x, &\mbox{for } u(t_k) = TR , &y = 0 \\
\lambda(1-\lambda)^{x+1-y}, &\mbox{for } u(t_k) = TR , &2 \leq y \leq x+1\\
(1-\lambda)^{x-1}, &\mbox{for } u(t_k) = NT, &y = 1 \\
\lambda(1-\lambda)^{x-y}, &\mbox{for } u(t_k) = NT, &2 \leq y \leq x
\end{array}
\right.
\label{system_equations}
\end{multline}
\fi
For $x \geq 1$, the reward function, which counts the number of time slots between successful transmissions by the adaptive user, is given by
\iftrue
\begin{multline}
\label{cost_function}
g\left(X(t_k)=x,u(t_k),X(t_{k+1})=y\right) \\ =
\left\{
\begin{array}{lll}
0, &\mbox{for $u(t_k) = TR$}, &\mbox{$y = 0$} \\
-2, &\mbox{for $u(t_k) = TR$}, &\mbox{$y \geq 2$} \\
-1, &\mbox{for $u(t_k) = NT$}, &\mbox{$y \geq 1$}
\end{array}
\right.
\end{multline}
\fi

We now explain \eqref{system_equations} and \eqref{cost_function}.  Suppose at stage $k$, $X(t_k) = x \geq 1$.  If the policy selects $u(t_k) = TR$, with probability $(1-\lambda)^x$ the uncooperative user's queue will be empty and the adaptive user will obtain a successful transmission.  Then, $X(t_{k+1}) = 0$ and no cost is incurred.  However, with probability $1-(1-\lambda)^x$ the uncooperative user's queue will be nonempty.  When this happens, a collision occurs at time $t_k$, the adaptive user elects to not transmit at the next time slot to allow the uncooperative user to successfully transmit the previously collided packet, and the next subsequence decision point $t_{k+1} = t_{k} + 2$.  Thus, a total of two time slots are lost by this event (cf., \eqref{cost_function} case 2) and a new value $X(t_{k+1})$ is obtained according to the distribution of \eqref{system_equations} case $2$.

If, on the other hand, the policy selects $u(t_k) = NT$ the adaptive user cannot obtain a successful transmission but also cannot incur a collision.  Then, the next subsequent decision point is $t_{k+1} = t_{k} + 1$, one time slot is lost (cf., \eqref{cost_function} case 3), and the next value of $X(t_{k+1})$ can be shown to be given by \eqref{system_equations} cases $3$ and $4$.

Given the transition probabilities \eqref{system_equations} and reward process \eqref{cost_function}, our objective is to obtain a policy $\pi = f_0, f_1, \dots$ that maximizes
\begin{equation}
\label{total_cost}
\sigma \triangleq \underset{K \to \infty}{\lim} \left. E\left[\sum_{k=0}^{K-1} g\left(X(t_k),u(t_k),X(t_{k+1})\right) \right| X(t_0) = 1\right].
\end{equation}
We define $\sigma^*$ to be the maximum, attainable value of \eqref{total_cost} over the set of all policies.

Note that following a successful adaptive user transmission, $W(t) = 1$.  Then, \eqref{total_cost} can be seen to be the (negative) expected number of time slots until the beginning of the next successful transmission by the adaptive user starting from a time slot immediately following a successful adaptive user transmission.  Using \eqref{elementary_renewal}, an upper bound on $\mu^*$ is
\begin{equation}
\label{upper_bound_equation_sigma}
\mu^*_{ub} \triangleq \frac{1}{1 - \sigma^*}.
\end{equation}

The specified Markov decision process is a negative expected total-reward problem with a countable state space and finite action space.  By \cite[Theorem~7.3.6]{puterman} such a problem has an optimal deterministic, stationary policy (i.e., there exists an optimal policy $\pi^* = f^*, f^*, \dots$ such that $f^*$ deterministically maps $f^*(X(t_k)) \in \mathcal{U}$).

Now, consider a deterministic stationary policy that chooses to transmit when $X(t_k) \in \left\{1, \dots, Y-1\right\}$ and not transmit when $X(t_k) = Y$ (i.e., the smallest numbered state in which the policy decides to not transmit is $Y$).  It follows that states $\{Y+1, Y+2, \dots \}$ are unreachable from states $\{0, 1, \dots, Y\}$.  Thus, states $\{0, 1, \dots, Y\}$ form a finite state Markov chain under this policy, and Bellman's equations for this policy have a solution over these states.  Denote by $V(x)$ the cost-to-go for state $x \in \{0,1,\dots,Y\}$. From equations \eqref{system_equations} and \eqref{cost_function} we see that Bellman's equations for the policy are given by the following.  For states $1,2,\dots,Y-1$ in which the policy transmits
\iftrue
\begin{multline}
V(x) = \sum_{y=1}^x \lambda(1-\lambda)^{x-y}\left(V(y+1) - 2 \right), \\ \mbox{ for } x = 1,2,\dots,Y-1.
\label{tx_bellman}
\end{multline}
\fi
For state $Y$ in which the policy does not transmit
\iftrue
\begin{equation}
V(Y) = (1-\lambda)^Y\left(V(1) - 1\right) + \sum_{y=1}^Y \lambda(1-\lambda)^{Y-y}\left( V(y) - 1 \right).
\label{idle_bellman}
\end{equation}
\fi
Furthermore, $V(0) = 0$, since once entering state $0$ no more negative reward is accrued.

By the definition of cost-to-go, $V(1)$ defined by \eqref{tx_bellman} is the same as \eqref{total_cost}, (i.e., $V(1)=\sigma$). For a given value of $\lambda$ we now optimize $V(1)$ over integer value $Y$ in order to characterize $\sigma^*$.  Note that if $Y = 1$, the controller chooses to never transmit and the expected time until a successful transmission is unbounded.  Thus, we restrict our attention to $Y \geq 2$.  The optimization is over the set of all policies that do not transmit in at least one state. However, it can be shown that as $Y$ goes to infinity, $V(1)$ approaches the value of $\sigma$ obtained by the policy that transmits in all states.


\subsubsection{Solving Bellman's Equations}

In this section we show that $\mu^*_{ub}$ can be found using a simple search over integer values $Y$.  We begin by giving the solution to $V(1)$ from Bellman's equations, \eqref{tx_bellman} and \eqref{idle_bellman}.
\begin{proposition}
\label{v1}
The cost-to-go of state $1$ of the stochastic shortest path problem is given by
\begin{equation}
\label{v1_eq}
V(1) = 
\left\{
\begin{array}{l}
\frac{\left(2 - 4\lambda \right)\left(\lambda (1-\lambda) \right)^{Y-1} + 2\lambda(1-\lambda)^{Y-1} - \lambda^{Y-1}}{(1-2\lambda)\left( \lambda^{Y-1} -  1\right)\left(1-\lambda\right)^{Y-1}}, \mbox{ for $\lambda \neq \frac{1}{2}$ } \\
\\
\frac{\left(\frac{1}{2}\right)^{Y-1} + Y - \frac{3}{2}}{\left(\frac{1}{2}\right)^Y - \frac{1}{2}}, \mbox{ for $\lambda = \frac{1}{2}$}
\end{array}
\right.
\end{equation}
\end{proposition}

We now give the following proposition for \eqref{v1_eq}.

\begin{proposition}
\label{maximum}
Over integers $Y>1$, there exists an $Y^*$ such that $V(1)$ is monotonically increasing for $1 < Y \leq Y^*$ and monotonically decreasing for $Y \geq Y^*$.
\end{proposition}

Proposition~\ref{maximum} implies that we can find $\mu^*_{ub}$ from \eqref{v1_eq} by a simple linear search over the integers $Y \geq 2$.  In Fig.~\ref{bounds_plot}, we plot $\mu^*_{ub}$ which was numerically found in this way.  We observe, $\mu^*_{ub}$ implies that policy $\pi^{lb}$ with $p=p^*$ performs well when the rate $\lambda$ is low. Note that it can be shown that there is a small gap between the bounds, even for small values of $\lambda$.  The gap goes to zero as $\lambda$ goes to zero.

\subsection{Stability of the Two User Network}
\label{stability_of_the_two_node_network}

We now briefly address the stability of the two user network.  Rigorous mathematical arguments for the following claims can be found in \cite[Chapter~3.2.4]{stahlbuhk_thesis}, but are omitted here for brevity.  Suppose packets arrive to the adaptive user's queue according to an \emph{i.i.d.} random process and wait to be successfully transmitted over the channel.  Then if the arrival rate is no greater than $\mu^*_{lb}$ it is easy to see that the above policy will rate stabilize both the adaptive and, by Corollary~\ref{primary_user_bounded_expectation}, uncooperative users' queues.  On the other hand, if the arrival rate is above $\mu^*_{ub}$, no policy exists that can stabilize the adaptive user.  Therefore, Fig.~\ref{bounds_plot} characterizes the throughput-stability region of the two user network.

\section{Networks with Multiple Users and Channels}
\label{section_multiple_user_network}

\subsection{Characterizing the Throughput-Stability Region}
\label{characterize_region}

We now examine networks with multiple adaptive and uncooperative users.  We begin by using the results of the previous section to define the stability region of larger networks.  To this end, for an uncooperative user assigned to channel resource $j$, with arrival rate $\lambda^j_u$, let $\mu^j_{lb}$ and $\mu^j_{ub}$ be the corresponding lower and upper bounds defined by \eqref{lower_bound}~and~\eqref{upper_bound_equation_sigma}, respectively.  Then, a necessary condition for the network to be mean rate stable is the existence of variables $\rho_{(i,j)} \in [0,1]$ for $i \in \{1, 2, \dots, N\}$ and $j \in J(i)$ such that
\begin{align}
& \sum_{j \in J(i)} \rho_{(i,j)} \geq \lambda^i_a, &\forall i \in \{1, 2, \dots, N\} \label{necessary_network_stable_1} \\
& \sum_{i:j \in J(i)} \rho_{(i,j)} \leq \mu^{j}_{ub}, & \forall j \in \{1, 2, \dots, M\}. \label{necessary_network_stable_2}
\end{align}
For a sufficient condition: if there exists variables $\rho_{(i,j)}$ such that 
\begin{align}
& \sum_{j \in J(i)} \rho_{(i,j)} \geq \lambda^i_a, &\forall i \in \{1, 2, \dots, N\} \label{sufficient_network_stable_1} \\
& \sum_{i:j \in J(i)} \rho_{(i,j)} \leq \mu^{j}_{lb}, & \forall j \in \{1, 2, \dots, M\} \label{sufficient_network_stable_2}
\end{align}
the network can be made rate stable (with probability one) by the following randomized policy, which we denote $\pi^R$. The controller uses implementations of policy $\pi^{lb}$ (of Subsection~\ref{lower_bound_section}) running independently on each channel to determine whether any adaptive user should transmit on that channel at the current time slot. Then, whenever policy $\pi^{lb}$ indicates a transmission attempt should occur, the controller chooses adaptive user $i$ to make the transmission with probability $\frac{\rho_{(i,j)}}{\sum_{i : j \in J(i)} \rho_{(i,j)}}$, where variables $\rho{(i,j)}$ are chosen to meet the above constraints. (If $\sum_{i : j \in J(i)} \rho_{(i,j)} = 0$, the channel is never scheduled.)  At each time slot, each adaptive user responds to the controller's schedule by transmitting a unique packet from its queue on each of its assigned channels, transmitting dummy packets on some channels if it is assigned more transmissions than it has queued packets.  


The above necessary and sufficient conditions follow from intuitive network flow constraints.  The complete proofs show stability by addressing in detail the limiting behavior of the queues under the above assumptions and can be found in \cite[Chapters~3.A.2~and~3.A.3]{stahlbuhk_thesis}. Specifically, the proof of the sufficient conditions uses \eqref{lower_bound_wp1} and that the randomized scheduling of adaptive users to channels converges to the average with probability one by the strong law of large numbers.

\subsection{Longest Queue First Scheduling}
\label{subsection_longest_queue_first} 

Policy $\pi^R$ uses the arrival rates $\lambda^i_a$ to make its scheduling decisions. In this subsection, we show that a policy, Longest Queue First (denoted $\pi^{LQF}$ in the following), that uses the adaptive users' queue backlogs to make scheduling decisions can achieve the same throughput-stability region as $\pi^R$.  The policy operates as follows.  The controller (again) uses implementations of policy $\pi^{lb}$ to determine whether any adaptive user should transmit on a channel.  Then, whenever policy $\pi^{lb}$ indicates a transmission attempt should occur on a channel, the controller chooses the adaptive user with the largest queue backlog that can transmit on the channel to make the transmission attempt (i.e., $\arg \max_{i:j \in J(i)} Q^i_a(t)$ with ties broken arbitrarily).  As above, each adaptive user responds to the schedule by transmitting a unique packet on each assigned channel, transmitting dummy packets when necessary.  We then have the following theorem that shows that LQF has the same throughput-stability region as $\pi^R$.

\begin{theorem}
\label{lqf_stabilizes_system_theorem}
If there exists variables $\rho_{(i,j)} \in [0, 1]$ satisfying conditions \eqref{sufficient_network_stable_1} and \eqref{sufficient_network_stable_2}, then $\pi^{LQF}$ achieves rate stability for all queues in the network.
\end{theorem}

Before giving the proof of the theorem, we would like to point out that if the arrival rates to all uncooperative users is no greater than $\frac{1}{3}$, $\pi^{lb}$ reduces to a policy that schedules an adaptive user transmission on a channel whenever a collision did not occur on that channel in the previous time slot.  Therefore, if the controller knows that the uncooperative users are greatly under-utilizing their resources, it can run LQF scheduling without any knowledge of the network's arrival rates.

\subsection{Throughput-Stability of Longest Queue First}

We now proceed to establish the proof to Theorem~\ref{lqf_stabilizes_system_theorem}.  Note that the proof of stability of the uncooperative users will follow from Corollary~\ref{primary_user_bounded_expectation}, and our focus will therfore be on the adaptive users.  Like the proof of MaxWeight \cite[Chapter 4.5]{georgiadis}, we will show  LQF's stability by comparing its performance to the randomized policy.  Unlike the original proof of MaxWeight, our proof will not rely on Lyapunov-drift methods, since we do not guarantee that the second moments of the service processes to the adaptive users are bounded.  Instead, the proof is based on analyzing sample path trajectories of policy $\pi^{LQF}$, a methodology originally used in \cite{armony} and \cite{ross}.

Our argument will proceed as follows.  Assume we are given an instantiation of the problem meeting the conditions of Theorem~\ref{lqf_stabilizes_system_theorem}. We will then analyze the performance of $\pi^{LQF}$ and $\pi^R$ using a sample path argument.  We let $\omega$ be an outcome of all randomness for our problem, including the arrival processes and randomness in the controller's policy.  Let $A^j_u(t, \omega)$ and $A^i_a(t, \omega)$ be the sample paths of the arrival processes.  Note that since both $\pi^R$ and $\pi^{LQF}$ use policy $\pi^{lb}$, for a fixed $\omega$, both policies when applied obtain successful transmissions on a channel at the same subset of time slots.  Thus, we denote the number of successful adaptive user transmissions obtained by both policies as $B^j(t, \omega) \in \{0, 1\}$.

Now, even though at each time $t$, under outcome $\omega$, both policies make the same decision as to whether an adaptive user should transmit on channel $j$, which adaptive user is assigned to make the transmission will vary between the two policies.  Thus, the sample paths of the random process $B^{(i,j)}_a(t)$ under polices $\pi^{LQF}$ and $\pi^{R}$ are different and are denoted $B^{(i,j),LQF}_a(t, \omega)$ and $B^{(i,j),R}_a(t, \omega)$ in the following.

Given the above, we are interested in analyzing the sample path trajectories, for time horizons $T = 1, 2, 3, \dots$.  Similar to the above, the sample path of random process $Q^i_a(t)$ under policies $\pi^{LQF}$ and $\pi^{R}$ are denoted $Q^{i,LQF}_a(t, \omega)$ and $Q^{i,R}_a(t, \omega)$, respectively.  We will now proceed to show that if for outcome $\omega$ there exists an adaptive user $i$ such that
\begin{equation}
\label{LQF_unstable}
\limsup_{T \to \infty} \frac{Q^{i,LQF}_a(T, \omega)}{T} > 0,
\end{equation}
then there must exist an adaptive user $i'$ not necessarily equal to $i$ such that
\begin{equation}
\label{randomized_unstable}
\limsup_{T \to \infty} \frac{Q^{i',R}_a(T, \omega)}{T} > 0.
\end{equation}
Since from Subsection~\ref{characterize_region} we know that with probability one the queues must be rate stable under $\pi^R$ for all arrival rates meeting conditions \eqref{sufficient_network_stable_1}~and~\eqref{sufficient_network_stable_2}, \eqref{randomized_unstable} does not occur with probability one and thus \eqref{LQF_unstable} does not occur.  Thus, the queues are rate stable under LQF if they are rate stable under the randomized policy.

At a high level, the proof will proceed as follows.  We will show that for any outcome $\omega$ such that \eqref{LQF_unstable} is true, there exists a set of  time slot intervals
\begin{equation*}
\left\{(a_1, b_1), (a_2, b_2), (a_3, b_3), \dots \right\}
\end{equation*} 
with linearly--growing durations, such that over each interval $(a_k, b_k)$ the LQF policy gave every transmission opportunity it could to a subset of adaptive users $I^* \subseteq \{1, 2, \dots, N\}$.  Despite this, we will show that at least one user in $I^*$ had a queue backlog that grew linearly over the intervals $(a_k, b_k)$.  Since over each interval, the randomized policy cannot give more service to $I^*$, one of its queue backlogs must also grow linearly under outcome $\omega$, implying \eqref{randomized_unstable} must also be true.

We begin with the following definition. For outcome $\omega$, define
\begin{equation*}
s(\omega) \triangleq \limsup_{T \to \infty} \sum_{i \in \{1, \dots, N\}} \frac{Q^{i,LQF}_a(T, \omega)}{T}.
\end{equation*}
Note that because the number of arrivals to any adaptive user at any given time is bounded above by $A_{max}$, $Q^{i,LQF}_a(T, \omega)$ is bounded by $A_{max} \times (T+1)$ for all $T = 1, 2, \dots$ (cf., \eqref{secondary_user_queue_size_summation}). It is then easy to see that $s(\omega) \in [0, N A_{max}]$.

We now establish two lemmas that will be instrumental in the proof of Theorem~\ref{lqf_stabilizes_system_theorem}.
\begin{lemma}
\label{first_lemma}
For any outcome $\omega$ such that for some $i \in \{1, 2, \dots, N\}$
\begin{equation}
\label{limsup_bigger_than_zero_assumption}
\limsup_{T \to \infty} \frac{Q^{i,LQF}_a(T, \omega)}{T} > 0,
\end{equation}
there exists a subset $I^* \subseteq \{1, 2, \dots, N\}$ and subsequences $Q^{i,LQF}_a(T_k, \omega)$ for $k = 1, 2, 3, \dots$ such that
\begin{align}
&\lim_{k \to \infty} \sum_{i \in I^*} \frac{Q^{i,LQF}_a(T_k, \omega)}{T_k} = s(\omega) \label{converges_to_s_I} \\
&\liminf_{k \to \infty} \frac{Q^{i,LQF}_a(T_k, \omega)}{T_k} > 0, &\forall i \in I^* \label{liminf_greater_than_0} \\
&\limsup_{k \to \infty} \frac{Q^{i,LQF}_a(T_k, \omega)}{T_k} = 0, &\forall i \not \in I^* \label{limsup_equal_to_0}.
\end{align}
\end{lemma}

The proof is in Appendix~\ref{appendix1}. Before continuing to the next lemma, we make the following definitions.
For a given subset $I^*$ and indices $T_k$ for $k = 1, 2, \dots$ satisfying \eqref{converges_to_s_I}, \eqref{liminf_greater_than_0}, and \eqref{limsup_equal_to_0}, define $\eta(\omega)$ to be a positive real number such that
\begin{equation}
\label{eta_lb_definition}
\eta(\omega) \leq \min_{i \in I^*} \left\{ \liminf_{k \to \infty} \frac{Q^{i,LQF}_a(T_k, \omega)}{T_k} \right\}.
\end{equation}
Note that $\eta(\omega)$ is a nonzero lower bound on the limit inferiors in \eqref{liminf_greater_than_0} and its admissible range is determined by the chosen subset $I^*$ and indices $T_k$.  Importantly, $\eta(\omega)$ cannot be greater than $A_{max}$, since $Q^{i,LQF}_a(T_k, \omega)$ cannot be greater than $A_{max}\times(T_k + 1)$.

Furthermore, define the set of uncooperative user channel resources $J(I^*)$ to be
\begin{equation*}
J(I^*) \triangleq \underset{i \in I^*}{\bigcup} J(i).
\end{equation*}

Now, consider an outcome $\omega$ such that \eqref{limsup_bigger_than_zero_assumption} is true, a subset $I^*$ and indices $T_k$ for $k = 1, 2, \dots$ satisfying \eqref{converges_to_s_I}, \eqref{liminf_greater_than_0}, and \eqref{limsup_equal_to_0}, and associated value $\eta(\omega)$ meeting \eqref{eta_lb_definition}.  For these parameters the following lemma holds.
\begin{lemma}
\label{interval_lemma}
For any $\epsilon \in \left(0, \frac{\eta(\omega)}{2}\right)$ and for all $k$ sufficiently large, over time interval $\left(T_k - S_k + 1, T_k - 1\right)$ with
\begin{equation}
\label{S_k_equation}
S_k = \left \lceil T_k \left( \frac{\eta(\omega) - 2 \epsilon}{M + A_{max}} \right) \right \rceil,
\end{equation}
the LQF policy only schedules adaptive users in $I^*$ to transmit on channels in $J(I^*)$ and no adaptive user in $I^*$ transmits dummy packets (i.e., its queue backlog is larger than the number of channels assigned to it).
\end{lemma}

The proof is in Appendix~\ref{appendix2}. We are now ready to establish the proof of the theorem.
\begin{IEEEproof}[Proof of Theorem~\ref{lqf_stabilizes_system_theorem}]
Consider an outcome $\omega$ such that \eqref{limsup_bigger_than_zero_assumption} is true for some $i$, a subset $I^*$ and indices $T_k$ for $k = 1, 2, \dots$ satisfying \eqref{converges_to_s_I}, \eqref{liminf_greater_than_0}, and \eqref{limsup_equal_to_0}, and associated value $\eta(\omega)$ meeting \eqref{eta_lb_definition}.  By Lemma~\ref{interval_lemma}, we know that for $k$ sufficiently large and $S_k$ given by \eqref{S_k_equation}, over the interval $(T_k - S_k + 1, T_k - 1)$ all adaptive users in $I^*$ do not transmit dummy packets and only the adaptive users in $I^*$ are scheduled to transmit on channels $J(I^*)$.  Defining $\tilde{B}^{i,LQF}(t, \omega)$ the number of packets to depart the queue of adaptive user $i$ at time $t$ for outcome $\omega$ under the LQF policy, this implies that for all $k$ sufficiently large
\iftrue
\begin{equation}
\label{we_get_all_departures}
\sum_{t =T_k - S_k + 1}^{T_k - 1} \sum_{i \in I^*} \tilde{B}^{i,LQF}_a(t, \omega)
= \sum_{t =T_k - S_k + 1}^{T_k - 1} \sum_{j \in J(I^*)} B^j(t, \omega).
\end{equation}
\fi
Now, for all $k$ sufficiently large, equation \eqref{secondary_user_queue_size_summation} implies
\iftrue
\begin{multline}
\label{update_equations_for_LQF_policy}
\sum_{i \in I^*} Q^{i,LQF}_a(T_k, \omega) - \sum_{i \in I^*} Q^{i,LQF}_a(T_k - S_k + 1, \omega) \\ = \sum_{T_k - S_k + 2}^{T_k} \sum_{i \in I^*} A^i_a(t, \omega) - \sum_{t =T_k - S_k + 1}^{T_k - 1} \sum_{i \in I^*} \tilde{B}^{i,LQF}_a(t, \omega).
\end{multline}
\fi
Applying equation \eqref{we_get_all_departures} to \eqref{update_equations_for_LQF_policy}, dividing by $S_k - 1$, and taking limits implies, by \cite[Theorem 3.19]{rudin}, that
\iftrue
\begin{align}
&\liminf_{k \to \infty} \frac{1}{S_k-1} \biggl( \sum_{i \in I^*} Q^{i,LQF}_a(T_k, \omega) \\
& \qquad \qquad \qquad \qquad \qquad - \sum_{i \in I^*} Q^{i,LQF}_a(T_k - S_k + 1, \omega) \biggr) \nonumber \\
&= \liminf_{k \to \infty} \frac{1}{S_k-1} \biggl( \sum_{T_k - S_k + 2}^{T_k} \sum_{i \in I^*} A^i_a(t, \omega)  \\
& \qquad \qquad \qquad \qquad \qquad - \sum_{t =T_k - S_k + 1}^{T_k - 1} \sum_{j \in J(I^*)} B^j(t, \omega) \biggr). \label{update_equations_for_LQF_policy_2}
\end{align}
\fi

We proceed to bound the left hand side of \eqref{update_equations_for_LQF_policy_2}.  We begin by noting that since $Q^{i,LQF}_a(T_k - S_k + 1, \omega)$ is a subsequence of $Q^{i,LQF}_a(T, \omega)$,
\begin{equation}
\label{back_end_limsup}
\limsup_{k \to \infty} \sum_{i \in I^*} \frac{Q^{i,LQF}_a(T_k - S_k + 1, \omega)}{T_k - S_k + 1} \leq s(\omega).
\end{equation}
Therefore, starting with the left hand side of \eqref{update_equations_for_LQF_policy_2},
\iftrue
\begin{align*}
& \liminf_{k \to \infty} \frac{1}{S_k-1} \biggl( \sum_{i \in I^*} Q^{i,LQF}_a(T_k, \omega) \\
& \qquad \qquad \qquad \qquad \qquad - \sum_{i \in I^*} Q^{i,LQF}_a(T_k - S_k + 1, \omega) \biggr) \\
& = \lim_{k \to \infty} \frac{T_k}{S_k-1} \sum_{i \in I^*} \frac{Q^{i,LQF}_a(T_k, \omega)}{T_k} \\
& \qquad \quad - \limsup_{k \to \infty} \frac{T_k - S_k + 1}{S_k-1} \sum_{i \in I^*} \frac{Q^{i,LQF}_a(T_k - S_k + 1, \omega)}{T_k - S_k + 1} \\
& \geq \left( \frac{M + A_{max}}{\eta(\omega) - 2 \epsilon} \right)s(\omega) - \left( \frac{M + A_{max}}{\eta(\omega) - 2 \epsilon} - 1 \right)s(\omega) \\
& = s(\omega)
\end{align*}
\fi
where the inequality follows from \eqref{converges_to_s_I} and \eqref{back_end_limsup} and the fact that
\begin{equation*}
\lim_{k \to \infty} \frac{T_k}{S_k - 1} = \frac{M + A_{max}}{\eta(\omega) - 2 \epsilon}.
\end{equation*}

Plugging this into \eqref{update_equations_for_LQF_policy_2},
\iftrue
\begin{multline}
\label{update_equations_for_LQF_policy_3}
s(\omega) \leq \liminf_{k \to \infty} \frac{1}{S_k-1} \biggl( \sum_{T_k - S_k + 2}^{T_k} \sum_{i \in I^*} A^i_a(t, \omega) \\ - \sum_{t =T_k - S_k + 1}^{T_k - 1} \sum_{j \in J(I^*)} B^j(t, \omega) \biggr).
\end{multline}
\fi

Now, define $\tilde{B}^{i,R}_a(t, \omega)$ to be the number of departures from the queue of adaptive user $i$ at time $t$ under the randomized policy for outcome $\omega$.  Then for outcome $\omega$, the randomized policy must have over interval $(T_k - S_k + 1, T_k-1)$ the following relationship
\begin{equation}
\label{randomized_can_only_do_worst}
\sum_{t =T_k - S_k + 1}^{T_k - 1} \sum_{i \in I^*} \tilde{B}^{i,R}_a(t, \omega) \leq \sum_{t =T_k - S_k + 1}^{T_k - 1} \sum_{j \in J(I^*)} B^j(t, \omega).
\end{equation}
This simply states that for outcome $\omega$, the number of departures from the adaptive user queues in set $I^*$ must be less than the total number of successful adaptive user transmissions (including dummy packets) over the channels in $J(I^*)$.
Moreover, the queue backlog of the adaptive user set $I^*$ at time $T_k$ for sample path $\omega$ under the randomized policy is bounded below by
\iftrue
\begin{multline}
\frac{1}{S_k - 1} \sum_{i \in I^*} Q^{i,R}_a(T_k, \omega)
\geq \frac{1}{S_k - 1} \biggl( \sum_{T_k - S_k + 2}^{T_k} \sum_{i \in I^*} A^i_a(t, \omega) \\ - \sum_{t =T_k - S_k + 1}^{T_k - 1} \sum_{i \in I^*} \tilde{B}^{i,R}_a(t, \omega) \biggr).
\label{queue_size_bounded_below_under_randomized}
\end{multline}
\fi
Combining \eqref{update_equations_for_LQF_policy_3}, \eqref{randomized_can_only_do_worst}, and \eqref{queue_size_bounded_below_under_randomized} we see that
\begin{align*}
s(\omega) &\leq \liminf_{k \to \infty} \sum_{i \in I^*} \frac{Q^{i,R}_a(T_k, \omega)}{S_k-1} \nonumber \\
 & = \left( \frac{M + A_{max}}{\eta(\omega) - 2 \epsilon} \right) \liminf_{k \to \infty} \sum_{i \in I^*} \frac{Q^{i,R}_a(T_k, \omega)}{T_k}. \nonumber
\end{align*}
This implies that 
\begin{equation*}
\limsup_{T \to \infty} \sum_{i \in I^*} \frac{Q^{i,R}_a(T, \omega)}{T} > 0.
\end{equation*}
Now, at any time $T$, 
\begin{equation*}
\max_{i \in I^*} \frac{Q^{i,R}_a(T, \omega)}{T} \geq \frac{1}{\left\lvert I^* \right\rvert} \sum_{i \in I^*} \frac{Q^{i,R}_a(T, \omega)}{T}.
\end{equation*}
Thus, there exists an $i' \in I^*$ such that 
\begin{equation*}
\limsup_{T \to \infty} \frac{Q^{i',R}_a(T, \omega)}{T} > 0.
\end{equation*}

Since, this argument holds for all $\omega$ and since for all $i \in \{1, 2, \dots, N\}$,
\begin{equation*}
P\left( \omega: \limsup_{T \to \infty} \frac{Q^{i,R}_a(T, \omega)}{T} > 0 \right) = 0,
\end{equation*}
this implies that
\begin{equation*}
P\left( \omega: \limsup_{T \to \infty} \frac{Q^{i,LQF}_a(T, \omega)}{T} > 0 \right) = 0
\end{equation*}
for all $i \in \{1, 2, \dots, N\}$, establishing the result.

We conclude by noting that since the LQF policy uses $\pi^{lb}$ to decide when to transmit on each channel $j$, by Corollary~\ref{primary_user_bounded_expectation}, the queues of the uncooperative users are stable.
\end{IEEEproof}

\subsection{Simulation}
\label{section_simulation}

\begin{figure*} 
    \centering
  \subfloat[\label{network_simulation_setup}]{%
       \includegraphics[width=0.37\linewidth]{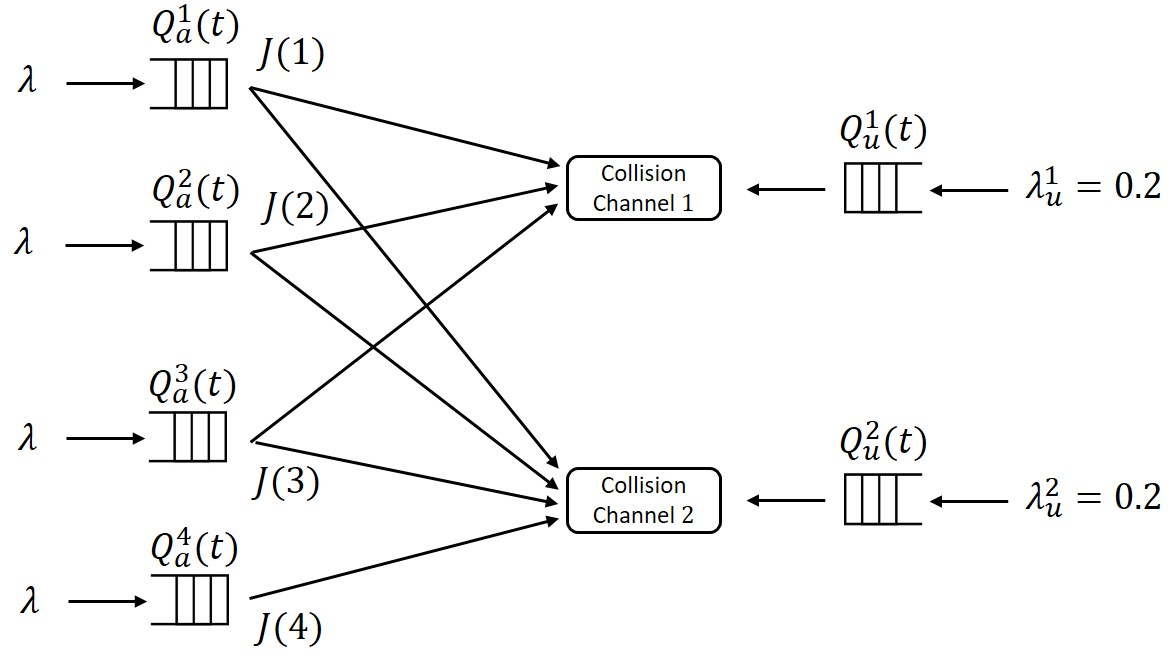}}
    \qquad \qquad \qquad
  \subfloat[\label{network_simulation_plot}]{%
        \includegraphics[width=0.35\linewidth]{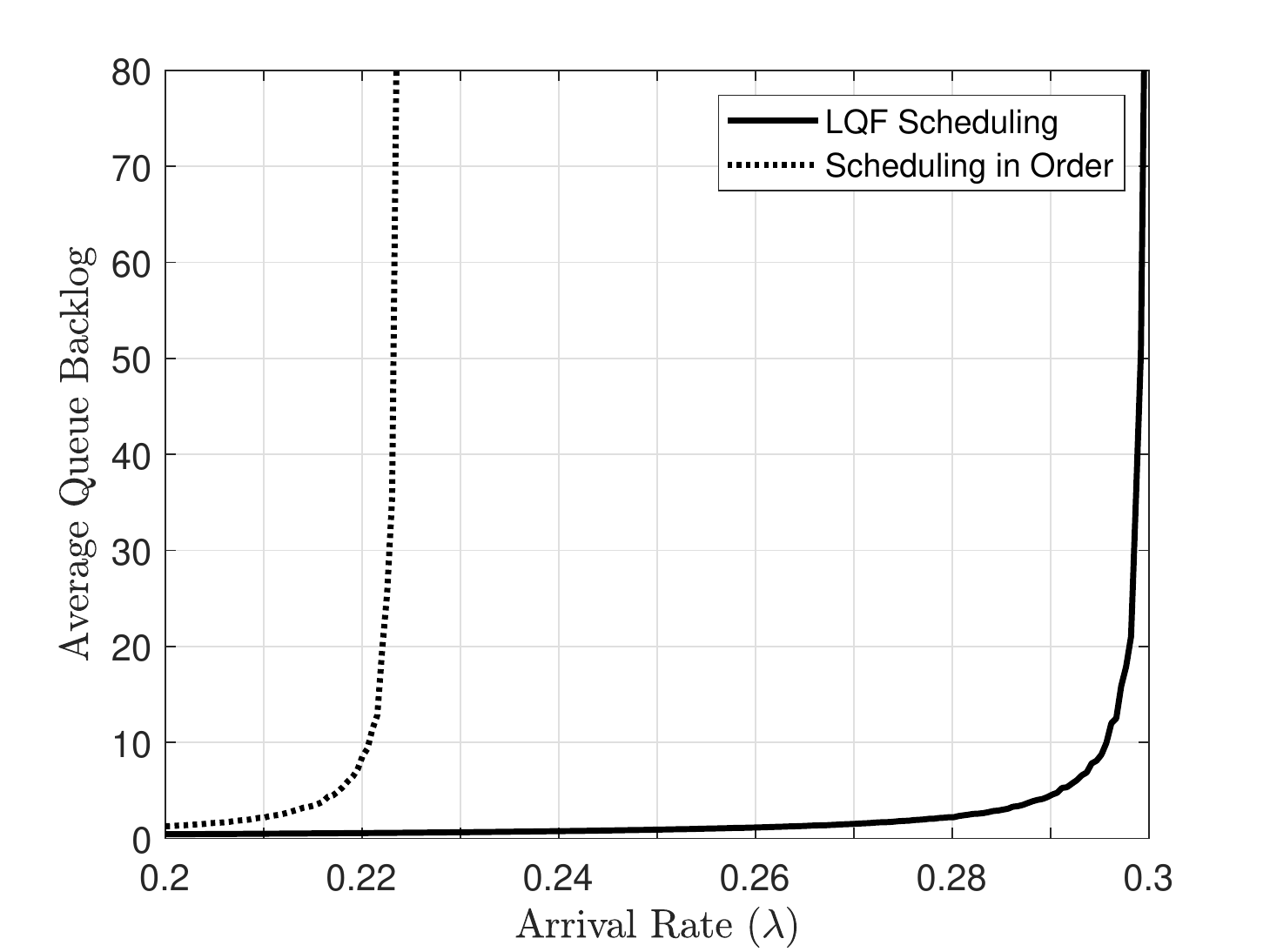}}
  \caption{(a) Illustration of the simulated network. (b) Average queue backlog for varying arrival rates to the adaptive users }
  \label{simulation_plots} 
\end{figure*}

In this section, we simulate the LQF policy on the network shown in Fig.~\ref{network_simulation_setup}.  In this network, there are four adaptive users and two uncooperative users assigned to two channels.  Three of the adaptive users can communicate on both channels, but the fourth can only transmit on channel $2$.  All adaptive users have the same arrival rate $\lambda$, and the uncooperative users have arrival rate $0.2$.  In Fig.~\ref{network_simulation_plot}, we show the average queue backlog in the network for varying arrival rates to the adaptive users (each simulated for 10 million time steps).  We note that under LQF scheduling, the backlogs are bounded up to an arrival rate of $0.3$, which is the edge of the sufficiency conditions given by \eqref{sufficient_network_stable_1}~and~\eqref{sufficient_network_stable_2}.

For contrast, we also show the average queue backlog for a policy that schedules the adaptive users in order from adaptive user $1$ to $4$; i.e., for each channel, the policy first schedules adaptive user $1$ to make transmissions if it has packets to send, if not it then goes to adaptive user $2$, etc.  Note that this policy is efficient at each time step. A channel is always scheduled to a user that has enough packets in its queue to use the channel resource.  In Fig.~\ref{network_simulation_plot}, we see that this policy becomes unstable well before LQF's boundary. The policy over allocates channel $2$ and leads to resource starvation, a problem avoided by LQF.  We therefore see that scheduling efficiently at each time slot is not enough to achieve good performance.  

\section{Assigning Uncooperative Users to Channel Resources}
\label{channel_assignment}

\begin{figure}[b!]
\centering
	\includegraphics[width=0.4 \textwidth]{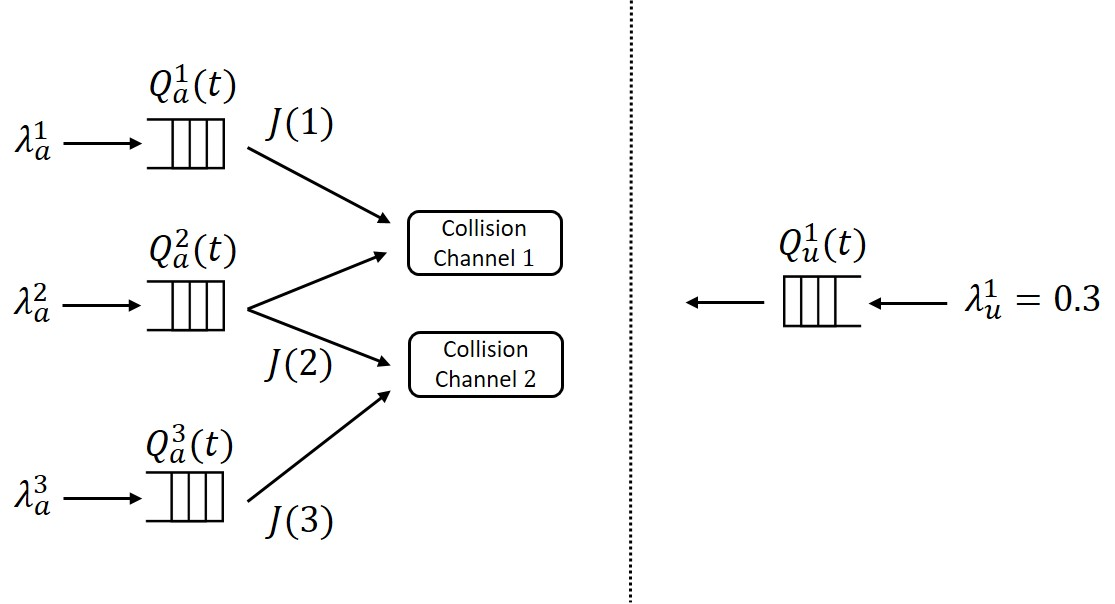}
\caption{The problem of assigning one uncooperative user to one of two channels.}
\label{assign0}
\end{figure}

\begin{figure*} 
    \centering
  \subfloat[]{%
        \includegraphics[width=0.4\linewidth]{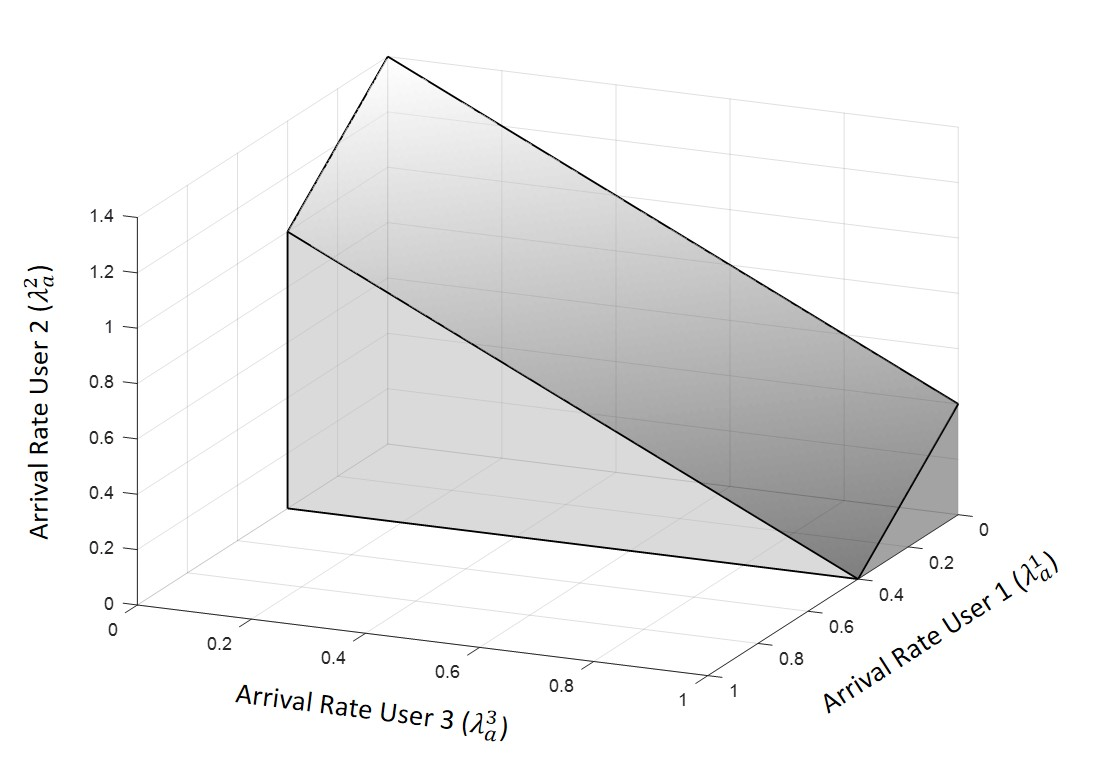}}
        \qquad \qquad \qquad
  \subfloat[]{%
        \includegraphics[width=0.4\linewidth]{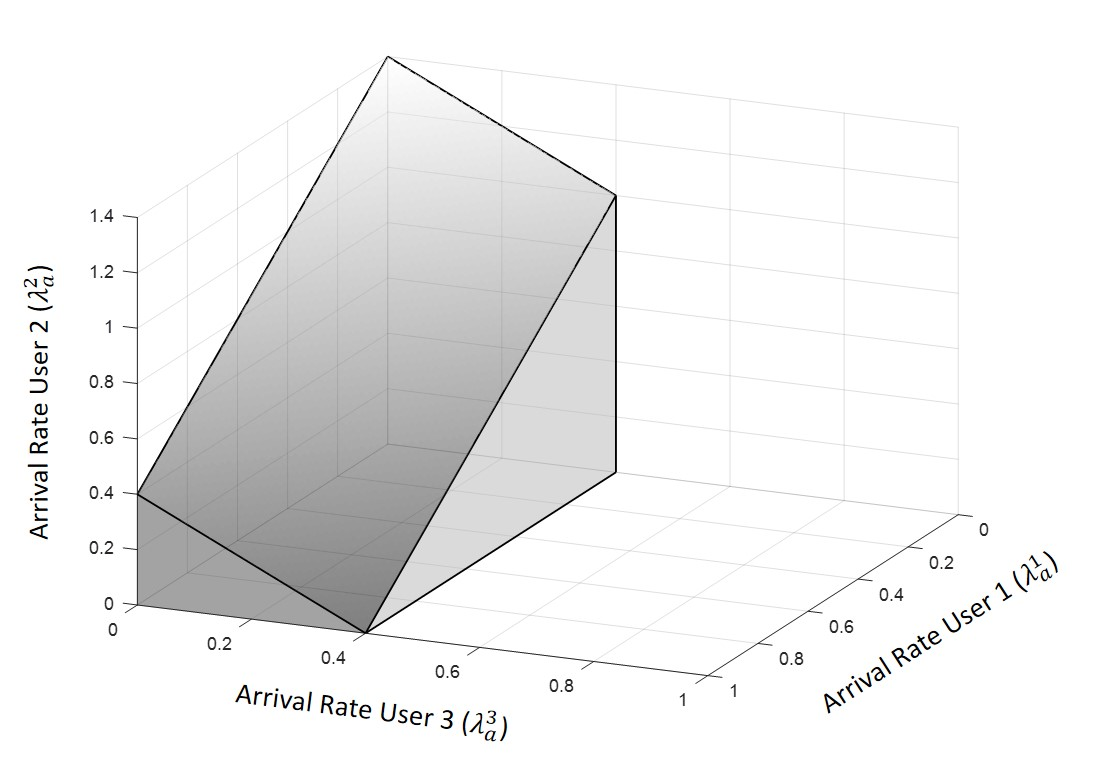}}
  \caption{(a) The resulting portion of the throughput-stability region that can be stabilized by $\pi^{LQF}$ if the uncooperative user in Fig.~\ref{assign0} is assigned to channel 1.  The region is only shown in the dimensions of the adaptive users' arrival rates for a fixed uncooperative user rate. (b) The region if assigned to channel 2.}
  \label{assign} 
\end{figure*}

In the previous sections, we assumed that the uncooperative users were preassigned to the channel resources by some unspecified process.  In some applications, this assignment may be done by an authority that is outside of the network controller's control.  However, in many other applications, although the uncooperative users may not be able to adaptively respond to a MaxWeight scheduler, the controller will still have the choice of which resource to assign them to at the start of time. For example, this problem could arise in a multiple access network, where some subset of nodes can only implement protocols that operate in a time division or frequency division mode.  These strict systems expect assignments to last for long time intervals, because reassignment of the channel resources would be laborious for the protocol.  After the assignment, the uncooperative users then use the assigned channels as described in the previous sections.  We now briefly extend our the results of the previous sections and address the problem of assigning uncooperative users to channel resources. In this section, we assume that each uncooperative user can only be assigned to one channel resource. Extending our framework to the case where uncooperative users can be assigned to multiple resources is straightforward.

We begin by noting that the assignment of uncooperative users to channels can greatly impact the throughput-stability region.  For example, in Fig.~\ref{assign0}, we show a network with three adaptive users and two channels, where two of the adaptive users can only transmit on one of the channels and one of the adaptive users can transmit on both.  Suppose we wish to assign one uncooperative user to one of the channels.  Clearly, the channel that is not chosen, advantages the adaptive users that transmit on that channel and shapes the throughput-stability region as shown in Fig.~\ref{assign}.  Note that neither region in Fig.~\ref{assign} is a subset of the other.

We now define the uncooperative user assignment problem.
\begin{definition}[Uncooperative User Assignment Problem]
Given a set of $N$ adaptive and $M$ uncooperative users with corresponding arrival rates, where each adaptive user $i$ can transmit on a subset of channels $J(i) \subseteq \{1, \dots, M\}$ and each uncooperative user $j$ can be assigned at the start of time to one channel resource in the subset $CH(j) \subseteq \{1, \dots, M\}$ (where at most one user can be assigned to a channel) does there exist an assignment of uncooperative users to channels such that the sufficient conditions of \eqref{sufficient_network_stable_1}~and~\eqref{sufficient_network_stable_2} are met?
\end{definition}
Or equivalently, is there an assignment such that the vector of arrival rates is within the resulting region that LQF can stabilize?  We show that this problem is NP-complete using a reduction from the set-covering decision problem.

\begin{theorem}
The uncooperative user assignment problem is NP-complete.
\end{theorem}
\begin{IEEEproof}
The set-covering decision problem is specified as follows.  Given a set of elements $E$ and a set of $S$ subsets, where each subset $s \subseteq E$ and the union of all subsets equals $E$, does there exist a choice of no more than $k$ subsets such that the union of the chosen subsets equals $E$?  Suppose we are given a set-covering problem.  Then, we construct a corresponding uncooperative user assignment problem as follows.  We add one adaptive user for each element in $E$ and give it an arrival rate equal to $\frac{1}{\lvert E \rvert}$.  For each subset in $S$, we also add one corresponding channel to our problem, and allow an adaptive user to transmit on the channel (i.e., $j \in J(i)$) if and only if its corresponding element is in that subset. We then add $\lvert S \rvert - k$ uncooperative users all with arrival rates equal to $1$ (always need to transmit) and $k$ uncooperative users with arrival rates $0$ (never need to transmit).  Any uncooperative user can be assigned to any channel.  It is then clear that the arrival rates can be supported if and only if we can assign the uncooperative users with arrival rates $0$ to the channels such that every adaptive user can transmit on at least one of those channels; this is equivalent to deciding if there is a choice of $k$ subsets such that each element is in at least one subset (i.e., the set-covering decision problem, which is NP-complete \cite[Chapter 35.3]{cormen}). An assignment of uncooperative users can be verified in polynomial time by using \eqref{sufficient_network_stable_1} and \eqref{sufficient_network_stable_2} to specify a simple linear program.  The result follows.
\end{IEEEproof}

For a given assignment of uncooperative users to channel resources, \eqref{sufficient_network_stable_1} and \eqref{sufficient_network_stable_2} define a sufficient condition for the resulting throughput-stability region.  When the controller can assign the channel resources to the uncooperative users, each assignment defines a different sufficient region, and one may view the union of these regions as the set of arrival rates that can be accommodated by the controller.  The above theorem shows that determining whether a given vector of arrival rates is within this union is in general NP-complete. Given an instance of the problem, any classical method for approaching mixed-integer linear programs could be used (e.g., simulated annealing, branch-and-bound, heuristics, etc.). See \cite[Chapter 11]{bertsimas} for further examples. Note that greedy algorithms have historically been a popular technique for set-covering problems, and are one approach that could be promising in our problem.  See \cite[Chapter 35.3]{cormen} for a description of how to apply greedy algorithms to the set-covering problem.

\section{Conclusion}
\label{section_conclusion}
In this work, we analyzed the throughput-stability regions of networks that have legacy users that are uncooperative with adaptive scheduling algorithms.  We showed that this region is shaped by how well the network controller can estimate the uncooperative users' backlogs and that the quality of the estimate is a function of the arrival rates to the uncooperative users.  We then determined that longest queue first scheduling combined with our estimation algorithm achieves a significant portion of the throughput-stability region, especially when the arrival rates to the uncooperative users are low.  We demonstrated the scheduling algorithm's performance in simulation and showed that it achieved queue stability up to the boundary given by our analysis.  We also proved that, in networks where the controller can assign uncooperative users to channel resources, assigning the uncooperative users to meet a desired traffic demand is an NP-complete problem.  A natural next-step is to design efficient heuristics to deal with this complexity, which is left for future work.

This work assumed independent arrivals to both the adaptive and uncooperative users.  This assumption is pessimistic, since we believe correlation between packet arrivals should allow the network controller to better estimate the uncooperative users' backlogs and attain larger throughput-stability regions.  With correlated packet arrivals, the controller would not only have to estimate the state of uncooperative users' queues but also the state of their arrival processes, which would add significant complexity.  Analyzing this problem domain is a direction for future work.


%

\appendices

\section{}
\label{appendix1}

\begin{IEEEproof}[Proof of Lemma~\ref{first_lemma}]
Assume that for outcome $\omega$ there exists at least one adaptive user $i$ such that
\begin{equation*}
\limsup_{T \to \infty} \frac{Q^{i,LQF}_a(T, \omega)}{T} > 0.
\end{equation*}
 Then, clearly $s(\omega) > 0$.  Now for each subset $I \subseteq \{1, \dots, N\}$ there must likewise exist a finite nonnegative value $s_I(\omega)$ such that
\begin{equation*}
\limsup_{T \to \infty} \sum_{i \in I} \frac{Q^{i,LQF}_a(T, \omega)}{T} = s_I(\omega).
\end{equation*} 
Note that for any two subsets $I$ and $I'$, if $I' \subseteq I$, $s_{I'}(\omega) \leq s_I(\omega)$.

Now, consider a subset $I^* \subseteq \{1, \dots, N\}$ such that: (1) $s_{I^*}(\omega) = s(\omega)$ and (2) there does not exist a strict subset $I' \subsetneq I^*$ such that $s_{I'}(\omega) = s(\omega)$. \footnote{For the empty set $\emptyset$, we define $s_\emptyset(\omega) \triangleq 0$.} For each $i \in I^*$, consider a subsequence of $Q^{i,LQF}_a(T, \omega)$ denoted $Q^{i,LQF}_a(T_k, \omega)$ for $k = 1, 2, 3, \dots$ such that \eqref{converges_to_s_I} holds.  By the definition of the limit superior \cite[Definition 3.16]{rudin} such a subsequence must exist.  Then over indices $T_k$, equations \eqref{liminf_greater_than_0} and \eqref{limsup_equal_to_0} must also hold.

To understand equation \eqref{liminf_greater_than_0}, note that if there existed a user $i^* \in I^*$ such that 
\begin{equation*}
\liminf_{k \to \infty} \frac{Q^{i^*,LQF}_a(T_k, \omega)}{T_k} = 0
\end{equation*}
we could construct a subsequence of $Q^{i,LQF}_a(T_k, \omega)$ denoted $Q^{i,LQF}_a(T_{k_\ell}, \omega)$ for $\ell \in \{1, 2, 3, \dots \}$ such that 
\begin{equation*}
\lim_{\ell \to \infty} \frac{Q^{i^*,LQF}_a(T_{k_\ell}, \omega)}{T_{k_\ell}} = 0.
\end{equation*}
But then, this would imply that for the subset $I' \triangleq I^* - i^*$,
\begin{align*}
&\lim_{\ell \to \infty} \sum_{i \in I'} \frac{Q^{i,LQF}_a(T_{k_\ell}, \omega)}{T_{k_\ell}} \nonumber \\
& = \lim_{\ell \to \infty} \sum_{i \in I^*} \frac{Q^{i,LQF}_a(T_{k_\ell}, \omega)}{T_{k_\ell}} - \lim_{\ell \to \infty} \frac{Q^{i^*,LQF}_a(T_{k_\ell}, \omega)}{T_{k_\ell}} \nonumber \\
& = s(\omega) - 0 \nonumber
\end{align*}
which implies 
\begin{equation*}
\limsup_{T \to \infty} \sum_{i \in I'} \frac{Q^{i,LQF}_a(T, \omega)}{T} = s(\omega),
\end{equation*}
contradicting the definition of $I^*$.

Likewise, equation \eqref{limsup_equal_to_0} may be seen by noting that if there existed a user $i' \not \in I^*$ such that 
\begin{equation*}
\limsup_{k \to \infty} \frac{Q^{i',LQF}_a(T_k, \omega)}{T_k} > 0,
\end{equation*}
then for the superset $I' \triangleq I^* \cup i'$,
\begin{align*}
&\limsup_{k \to \infty} \sum_{i \in I'} \frac{Q^{i,LQF}_a(T_k, \omega)}{T_k} \nonumber \\
& = \lim_{k \to \infty} \sum_{i \in I^*} \frac{Q^{i,LQF}_a(T_k, \omega)}{T_k} + \limsup_{k \to \infty} \frac{Q^{i',LQF}_a(T_k, \omega)}{T_k} \nonumber \\
& > s(\omega). \nonumber
\end{align*}
But, this would imply 
\begin{equation*}
\limsup_{T \to \infty} \sum_{i \in I'} \frac{Q^{i,LQF}_a(T, \omega)}{T} > s(\omega)
\end{equation*}
which contradicts the definition of $s(\omega)$.
\end{IEEEproof}

\section{}
\label{appendix2}

\begin{IEEEproof}[Proof of Lemma~\ref{interval_lemma}]
Equations \eqref{liminf_greater_than_0} and \eqref{limsup_equal_to_0} imply by \cite[Theorem 3.17]{rudin} that for every $\epsilon > 0$, there exists a $k_0(\epsilon, \omega)$ such that for all $k \geq k_0(\epsilon, \omega)$
\begin{align}
&\frac{Q^{i,LQF}_a(T_k, \omega)}{T_k} \geq \eta(\omega) - \epsilon, &\forall i \in I^* \label{epsilon_conditions_3} \\
&\frac{Q^{i,LQF}_a(T_k, \omega)}{T_k} \leq \epsilon, &\forall i \not \in I^* \label{epsilon_conditions_4}
\end{align}
where $\eta(\omega)$ satisfies \eqref{eta_lb_definition}.

For some choice of $\epsilon < \frac{\eta(\omega)}{2}$ consider some time $T_k$ for $k \geq k_0(\epsilon, \omega)$.  We wish to bound the last time we could have scheduled an adaptive user that was not in $I^*$ to communicate on a channel in the set $J(I^*)$.  Under LQF scheduling, the last time this could have occurred was when a user not in $I^*$ had a larger queue than a user in $I^*$.  Since the rate at which packets may leave any adaptive user queue is upper bounded by $M$ and the rate that packets arrive at any adaptive user is upper bounded by $A_{max}$, we see that from \eqref{epsilon_conditions_3} and \eqref{epsilon_conditions_4}, the minimum length of time since the last time we scheduled a user not in $I^*$ to transmit over a channel in $J(I^*)$, denoted $S_k$, is given by 
\begin{equation*}
S_k \triangleq \min \left \{S \in \mathbbm{Z}_+ : T_k(\eta(\omega) - \epsilon) - S A_{max} \leq T_k \epsilon + S M \right\}.
\end{equation*}
 Solving for $S_k$ gives \eqref{S_k_equation}.

Now, consider the interval $\left( T_k - S_k + 1, T_k - 1 \right)$.  \footnote{For $T_k \geq 2\left( \frac{M + A_{max}}{\eta(\omega) - 2 \epsilon} \right)$, $S_k \geq 2$, and the interval defined by $\left( T_k - S_k + 1, T_k - 1 \right)$ is nonempty.  Without loss of generality, we assume this is true for all $k$.}  Over this interval we exclusively scheduled users $i \in I^*$ to communicate on channels $J(I^*)$.  Moreover, if for all $k = 1, 2, \dots$,\footnote{This condition guarantees that $T_k - S_k + 1 < T_{k+1} - S_{k+1} + 1$ with $S_k$ and $S_{k+1}$  defined by \eqref{S_k_equation}.}
\begin{equation*}
T_{k+1} - T_k > \frac{1}{1 - \frac{\eta(\omega) - 2\epsilon}{M + A_{max}}},
\end{equation*}
then the sequence $Q^{i,LQF}_a(T_k - S_k + 1, \omega)$ is a subsequence of $Q^{i,LQF}_a(T, \omega)$ (i.e., $T_k - S_k + 1$ is unique for each $k$ and increasing in $k$).  Without loss of generality, we assume this condition is met.

Now, an adaptive user may transmit dummy packets only when it is scheduled to transmit on more channels than it has packets in queue.  Thus, as a sufficient condition, if the backlog is greater than $M$ at time $t$, we may be certain that no dummy packets are transmitted at time $t$.  We thus lower bound the interval of time preceding $T_k$ since an adaptive user $i \in I^*$ has had a queue backlog less than $M$.    We denote the bound $\hat{S}_k$.    By an argument similar to the above, one may see this is given by 
\begin{equation*}
\hat{S}_k \triangleq \min \left \{S \in \mathbbm{Z}_+ : T_k(\eta(\omega) - \epsilon) - S A_{max} \leq M \right\},
\end{equation*}
 which implies 
\begin{equation*}
\hat{S}_k = \max \left\{0, \left \lceil T_k\left( \frac{\eta(\omega) - \epsilon}{A_{max}}\right) - \frac{M}{A_{max}}   \right \rceil \right\}.
\end{equation*}
Thus, for $T_k$ sufficiently large, $S_k \leq \hat{S}_k$, and we see that over the interval $(T_k - S_k + 1, T_k - 1)$ the adaptive users in $I^*$ do not transmit dummy packets.
\end{IEEEproof}



\ifCLASSOPTIONcaptionsoff
  \newpage
\fi



%

%

\begin{IEEEbiography}[{\includegraphics[width=1in,height=1.25in,clip,keepaspectratio]{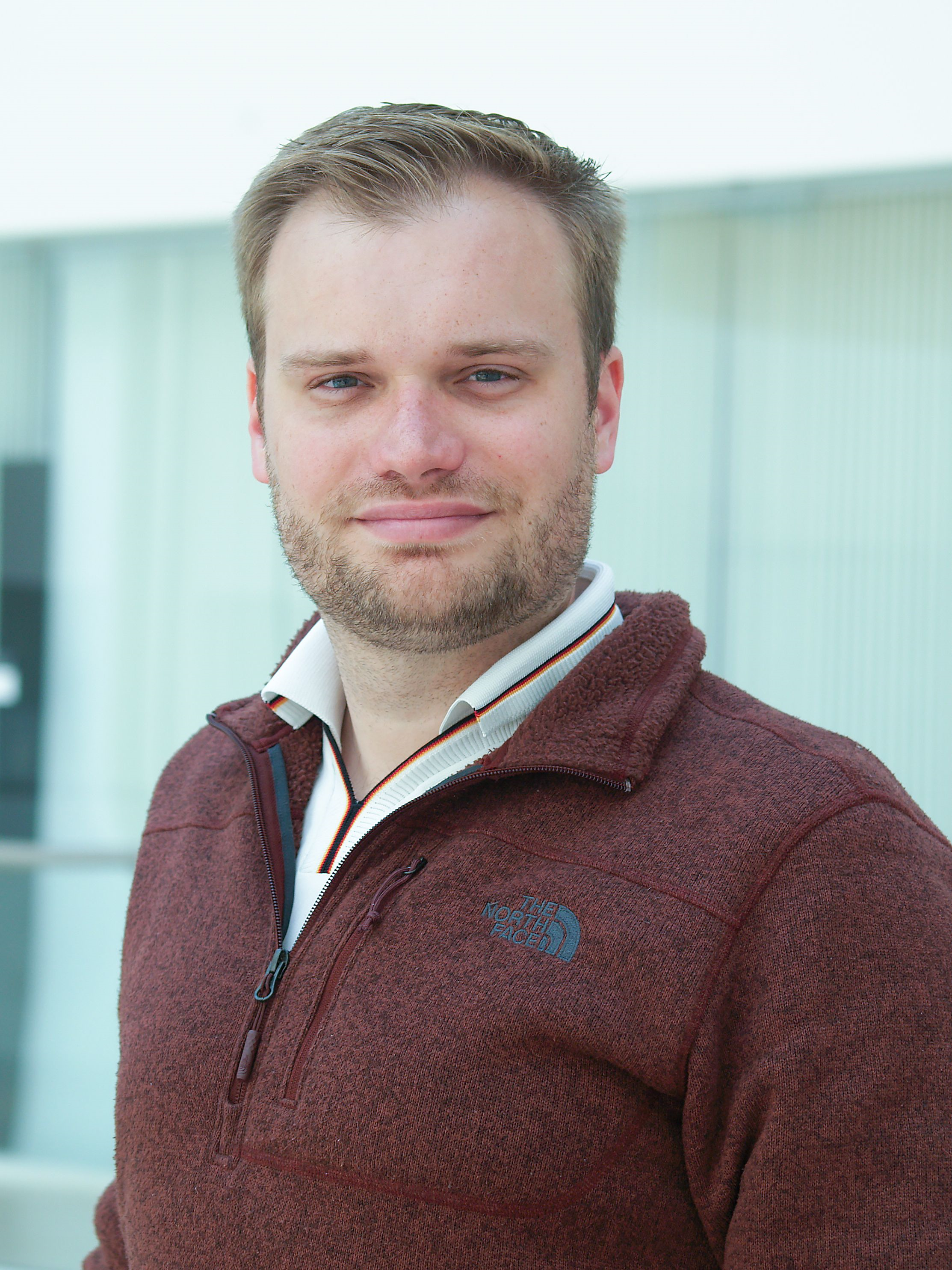}}]{Thomas Stahlbuhk}
received his B.S. and M.S. degrees in Electrical Engineering from the University of California San Diego in 2008 and 2009, respectively, and received his Ph.D. degree in Communications and Networks from the Massachusetts Institute of Technology (MIT) in 2018. He is currently a member of the Technical Staff at MIT Lincoln Laboratory, working in communication networks research. His research interests are in wireless networks, optimal control, and applied probability.
\end{IEEEbiography}

\begin{IEEEbiography}[{\includegraphics[width=1in,height=1.25in,clip,keepaspectratio]{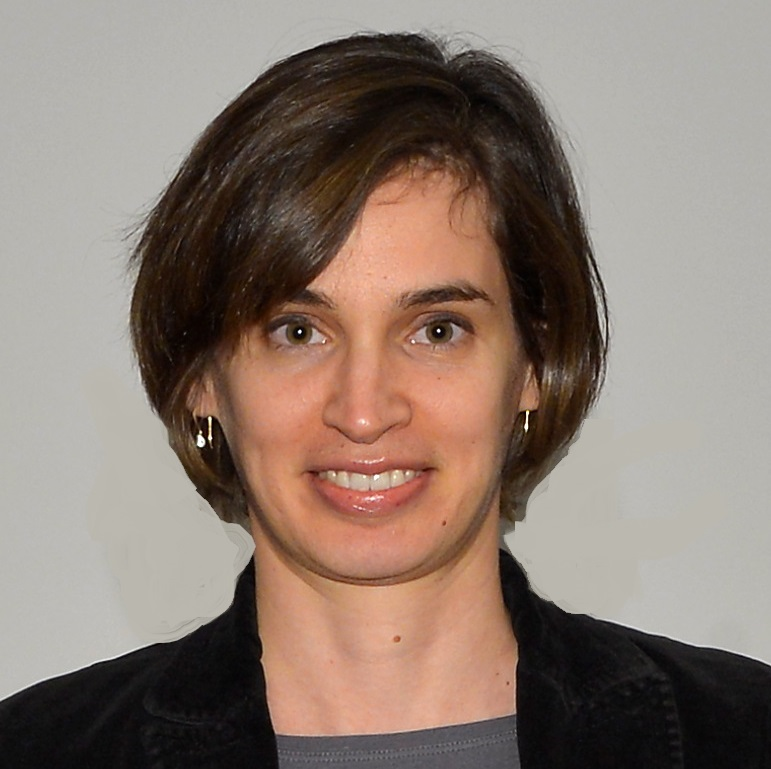}}]{Brooke Shrader}
received the B.S. degree from Rice University, the M.S. degree from the Swedish Royal Institute of Technology (KTH), and the Ph.D. degree from the University of Maryland, College Park, all in electrical engineering. She is a Senior Member of Technical Staff with the Massachusetts Institute of Technology Lincoln Laboratory, where she has been since 2008. Her research interests lie in communication systems, wireless networks, and related disciplines, including information theory, control, and queueing models. She currently serves as Associate Editor for the IEEE/ACM Transactions on Networking.
\end{IEEEbiography}

\begin{IEEEbiography}[{\includegraphics[width=1in,height=1.25in,clip,keepaspectratio]{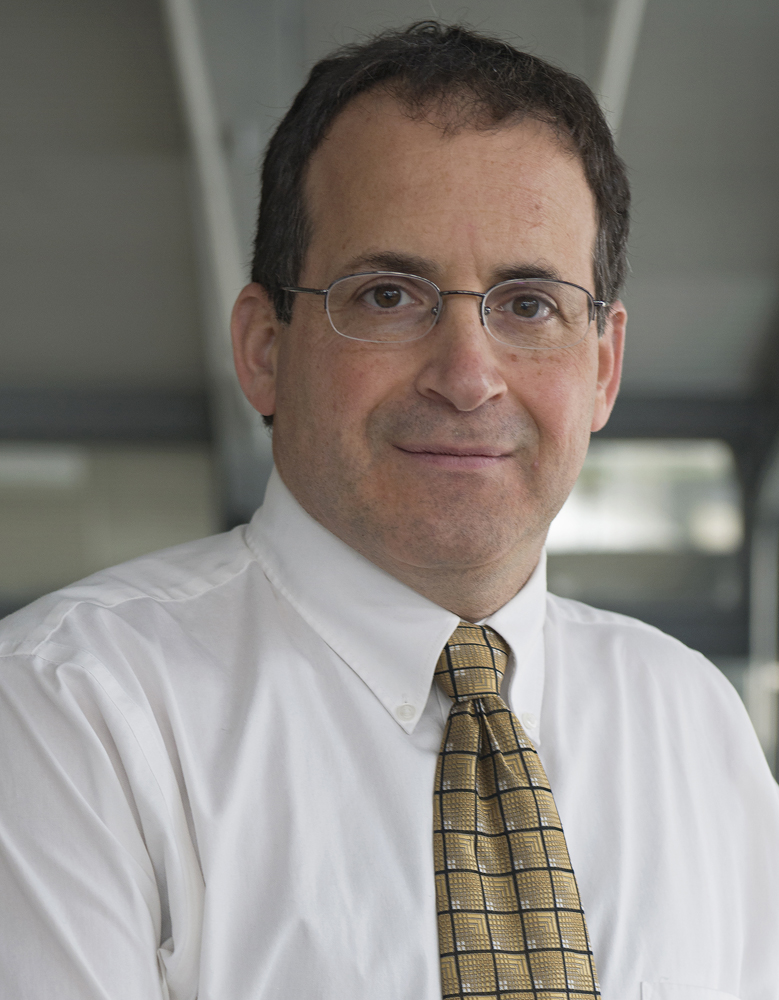}}]{Eytan Modiano}
is Professor in the Department of Aeronautics and Astronautics and Associate Director of the Laboratory for Information and Decision Systems (LIDS) at MIT.  Prior to Joining the faculty at MIT in 1999, he was a Naval Research Laboratory Fellow between 1987 and 1992, a National Research Council Post Doctoral Fellow during 1992-1993, and a member of the technical staff at  MIT Lincoln Laboratory between 1993 and 1999.  Eytan Modiano received his B.S. degree in Electrical Engineering and Computer Science from the University of Connecticut at Storrs in 1986 and his M.S. and PhD degrees, both in Electrical Engineering, from the University of Maryland, College Park, MD, in 1989 and 1992 respectively.

His research is on modeling, analysis and design of communication networks and protocols.  He received the Infocom Achievement Award (2020) for contributions to the analysis and design of cross-layer resource allocation algorithms for wireless, optical, and satellite networks.   He is the co-recipient of the Infocom 2018 Best paper award, the MobiHoc 2018 best paper award, the MobiHoc 2016 best paper award, the Wiopt 2013 best paper award, and the Sigmetrics 2006 best paper award.  He was the Editor-in-Chief for IEEE/ACM Transactions on Networking (2017-2020), and served as Associate Editor for IEEE Transactions on Information Theory and IEEE/ACM Transactions on Networking.  He was the Technical Program co-chair for  IEEE Wiopt 2006, IEEE Infocom 2007, ACM MobiHoc 2007, and DRCN 2015.  He had served on the IEEE Fellows committee in 2014 and 2015, and is a Fellow of the IEEE and an Associate Fellow of the AIAA.
\end{IEEEbiography}





\end{document}